\documentclass[fleqn,usenatbib]{mnras}

\usepackage{newtxtext,newtxmath}
\usepackage[T1]{fontenc}

\DeclareRobustCommand{\VAN}[3]{#2}
\let\VANthebibliography\thebibliography
\def\thebibliography{\DeclareRobustCommand{\VAN}[3]{##3}\VANthebibliography}

\usepackage{graphicx}	% Including figure files
\usepackage{amsmath}	% Advanced maths commands
\usepackage{color}
\usepackage{rotating}
\usepackage{pdflscape}
\usepackage[export]{adjustbox}
\usepackage{hyperref}
\usepackage{CJKutf8}

\defcitealias{macfarlane21}{M21}
\title[Decomposing quasar radio emission]{A novel Bayesian approach for decomposing the radio emission of quasars: I. Modelling the radio excess in red quasars}

\author[B.-H. Yue et al.]{
B.-H. Yue~\begin{CJK*}{UTF8}{gbsn}(岳博涵)\end{CJK*},$^{1}$\thanks{E-mail: bohan.yue@ed.ac.uk (BY)}
P. N. Best,$^{1}$
K. J. Duncan,$^{1}$
G. Calistro-Rivera,$^{2}$
L. K. Morabito,$^{3,4}$
J. W. Petley,$^{3}$
\newauthor
I. Prandoni,$^{5}$
H. J. A. R\"ottgering,$^{6}$
D. J. B. Smith$^{7}$
\\
$^{1}$Institute for Astronomy, University of Edinburgh, Edinburgh EH9 3HJ, UK\\
$^{2}$European Southern Observatory, Karl-Schwarzchild-Str. 2, 85748, Garching bei M\"unchen, Germany\\
$^{3}$Centre for Extragalactic Astronomy, Department of Physics, Durham University, Durham DH1 3LE, UK\\
$^{4}$Institute for Computational Cosmology, Department of Physics, University of Durham, South Road, Durham DH1 3LE, UK\\
$^{5}$INAF-IRA, Via P. Gobetti 101, 40129 Bologna, Italy\\
$^{6}$Leiden Observatory, Leiden University, PO Box 9513, NL-2300 RA Leiden, The Netherlands\\
$^{7}$Centre for Astrophysics Research, University of Hertfordshire, College Lane, Hatfield AL10 9AB, UK\\
}

\date{Accepted XXX. Received YYY; in original form ZZZ}

\pubyear{2024}

\begin{document}
\label{firstpage}
\pagerange{\pageref{firstpage}--\pageref{lastpage}}
\maketitle

\begin{abstract}
Studies show that both radio jets from the active galactic nuclei (AGN) and the star formation (SF) activity in quasar host galaxies contribute to the quasar radio emission; yet their relative contributions across the population remain unclear. Here, we present an improved parametric model that allows us to statistically separate the SF and AGN components in observed quasar radio flux density distributions, and investigate how their relative contributions evolve with AGN bolometric luminosity ($L_\mathrm{bol}$) and redshift ($z$) using a fully Bayesian method. Based on the newest data from LOFAR Two-Metre Sky Survey Data Release 2, our model gives robust fitting results out to $z\sim4$, showing a quasar host galaxy SFR evolution that increases with bolometric luminosity and with redshift out to $z\sim4$. This differs from the global cosmic SFR density, perhaps due to the importance of galaxy mergers. The prevalence of radio AGN emissions increases with quasar luminosity, but has little dependence on redshift. Furthermore, our new methodology and large sample size allow us to subdivide our dataset to investigate the role of other parameters. Specifically, in this paper, we explore quasar colour and demonstrate that the radio excess in red quasars is due to an enhancement in AGN-related emission, since the host galaxy SF contribution to the total radio emission is independent of quasar colour. We also find evidence that this radio enhancement occurs mostly in quasars with weak or intermediate radio power.
\end{abstract}


%%%%%%%%%%%%%%%%%%%%%%%%%%%%%%%%%%%%%%%%%%%%%%%%%%

%%%%%%%%%%%%%%%%% BODY OF PAPER %%%%%%%%%%%%%%%%%%

\section{Introduction}
\label{sec:intro}

Quasars (QSOs) are known to have a profound impact on the evolution of their host galaxies, both through radiation-driven processes and through jet feedback \citep[e.g. see reviews by][]{fabian12, heckman14}. These physical processes and their impact on the host galaxies can be traced across multiple wavelengths, therefore helping us build knowledge towards the evolutionary paths of galaxies \citep[e.g.][]{kormendy13}{}{}. \par

Radio observations play a particularly important role in analysing the impact of active galactic nuclei (AGNs) on galaxy evolution, since they are a major tracer of AGN jets \citep[][]{heckman14,hardcastle20} {} {}, in which relativistic electrons create synchrotron radiation that is detectable in radio bands. AGN jets have a highly collimated structure originating from an optically thick launching region also known as the jet base \citep{blandford79,reynolds82}, and give rise to radio lobes with a steep continuum profile due to highly relativistic electrons. The non-thermal processes from relativistic electrons give the radio cores a high brightness temperature \citep{blundell98}, and radio emission from a jet origin is often highly polarized. These properties can be used to identify radio jets even in the lowest power regime, given sufficient angular resolution \citep[e.g. see the resolved detection of NGC 4151 in][]{pedlar93, williams17, mundell03}. \par

Radio emission can also arise from regions of star formation \citep[SF; e.g.][]{condon92}{}{}, which could be associated with the AGN activity. Radio emission from SF is characterised by its steep spectrum at low frequencies (with a spectral index\footnote{In this work the spectral index is defined as $\alpha$ assuming $L_\nu \propto\nu^{-\alpha}$.} of $\alpha\sim0.7$), which is caused by the acceleration of electrons in supernova remnants. \par

Separating SF and AGN features in radio-faint quasars is particularly difficult, as the hosts of most AGNs are star-forming galaxies \citep{heckman14}. The most popular approach where suitable multiwavelength optical / IR data are available is through spectral energy distribution (SED) fitting \citep[e.g.][]{rivera17,delvecchio17,whittam22,best23}{}{}; however, the number of sources with reliable SF-AGN identification is limited by the scope of optical/IR surveys and the number of available photometric bands. Recently, without having to utilize the multiwavelength data,  \citet{morabito22} applied very-long baseline interferometry (VLBI) techniques to calculate the radio brightness temperature and separate SF and AGN contributions in spatially-unresolved quasars, but similar tasks are still difficult to accomplish for sources without VLBI-resolution observations. Our lack of knowledge about the ongoing physical processes driving the radio emission in AGNs and their host galaxies has a profound impact in addressing some of the most important questions regarding the quasar properties and the impact of the quasars on their host galaxies. \par

Historically, quasars have been separated into two categories -- radio-quiet (RQ) quasars and radio-loud (RL) quasars -- based on their radio-loudness R, which is typically defined as the ratio between fluxes in the optical and radio bands: $R=f(4400\mathring{A})/f(6\mathrm{cm})$ \citep{kellerman89}. Results from large radio surveys indicated an apparent dichotomy between RL and RQ quasars, thanks to an asymmetric distribution of radio flux densities with a long tail towards the radio-bright end (due to powerful jet activities), combining with a peak at low flux density regime (due to host galaxy star formation or small-scale radio emissions from the AGN). However, debates are still open on whether star formation in host galaxies can provide sufficient radio emission as observed in RQ quasar samples \citep[for a thorough review on the topic, see][]{panessa19}: some studies found that the radio emission in RQ quasars can be explained by SF alone \citep[e.g.][]{kimball11,condon13,bonzini13}, while others suggest that the majority of the emission needs to come from AGN activity \citep[e.g.][]{zakamska16, white15, white17}, in the form of small-scale jets, AGN winds \citep{mullaney13,zakamska16,morabito19,petley22}, or accretion disk coronae \citep{laor08,chen23}. In support of the latter argument which links RQ quasar radio emission with AGN activities, weak radio jets have recently been identified within several spatially resolved RQ quasars \citep[e.g.][]{leipski06,herrera16,jarvis19}. These studies suggest that the weak jet activity in RQ AGNs might be merely a scaled-down version of RL jets, and the only difference lies in the powering efficiency of acceleration on sub-parsec scales, which therefore would indicate that radio jets may contribute to the radio emission in \emph{all} quasars, whether they are low-luminosity small-scaled or radio-loud extended. \par

The LOw-Frequency ARray \citep[LOFAR;][]{lofar} is a state-of-art radio telescope observing at 120-168 MHz with its high-band antennae. Thanks to LOFAR's wide field of view combined with its overall sensitivity, including sensitivity to low surface brightness emission, we can now detect radio sources at an order of magnitude higher sky density than any previous large-area radio surveys and obtain deep radio images for a large sample of quasars with optical counterparts (see Section~\ref{sec:data_lotss} for details). 

Motivated by the high sensitivity of LOFAR observations, \citet[][hereafter \citetalias{macfarlane21}]{macfarlane21}{}{} studied in detail the radio flux density distribution of quasars from the Sloan Digital Sky Survey \citep[SDSS; e.g.][]{dr14q}. They adopted the conclusion from \citet{gurkan19} that every source hosts a contribution from both jet activity in the AGN and star formation in the host galaxy, and proposed a two-component model that characterises the radio emission in the quasar population, where both the SF and AGN components were modelled from physical descriptions. The contributions to the overall quasar radio emission from host galaxy SF and AGN jets can thus be studied independently with the two-component model. While investigating the variation of radio emission from SF and jet components with bolometric luminosity, redshift, and black hole (BH) mass, their two-component model was able to provide a good fit to the data across all parameter space, thus strongly indicating the lack of an RL/RQ dichotomy. \par

As more evidence point toward a continuous distribution of quasar jet power rather than a dichotomy \citep[e.g.][]{cirasuolo03a,cirasuolo03b,balokovic12}{}{}, which factors affect the powering efficiency of radio jets has thus become a more interesting topic for discussion. Some studies have argued that jet strengths are related to the bolometric luminosity of the AGN; earlier studies show a higher jet fraction in low-luminosity RQ quasars \citep[e.g.][]{ho01,blundell01,ulvestad01}, while in the \citetalias{macfarlane21} model the fraction of sources with powerful jets (denoted as $f$) does increase with $L_\mathrm{bol}$ sublinearly ($f\propto L_\mathrm{bol}^{0.65}$). Others have found strong dependencies between jet power and other parameters such as black hole mass \citep[e.g.][]{laor00,lacy01,mclure04,best05}. Some recent works \citep[including][]{retana17}{}{} have found higher angular clustering in RL quasars, suggesting that larger-scale environment might also be a factor. \citet{morabito19} found a lower RL fraction in broad-absorption line quasars (BALQSOs) which links the radio-loudness to an outflow phase. Based on their two-component model that separates the host galaxy contribution from the observed quasar radio emission, \citetalias{macfarlane21} has revealed positive correlations between typical jet power and optical luminosity or black hole mass — but not with redshift. Therefore, they suggested the production of radio jets is more likely to be governed by intrinsic properties. \par

Recently, it has been shown that a small population of unusually red quasars \citep[rQSOs; e.g.][]{richards03}{}{} shows a significant excess in radio emission compared to a control sample of QSOs with blue or average colours \citep[e.g.][]{glikman07,urrutia09,glikman12,klindt19,rivera21} {} {}. Several theories have been proposed to explain the nature of this rQSO population. \citet[][and references therein]{netzer15} argued that from the standard AGN model's point of view, the red quasars are simply typical blue quasars with their core area (the accretion disc and broad line region) partially obscured by the dusty torus. However, more recent studies have connected red quasars with other phenomena including flatter bolometric luminosity functions \citep[e.g.][]{banerji15}, occurrence of major mergers \citep[e.g.][]{urrutia08, glikman15} and higher incidence of strong AGN outflows \citep[e.g.][]{urrutia09,banerji12}. These phenomena cannot be explained by the torus obscuration \citep[see][]{rivera21}. More specifically, based on the higher prevalence of radio activity in the SDSS-selected red quasar sample, \citet{klindt19} concluded that red quasars are a fundamentally different population from the typical blue quasars \citep[see also][]{rosario20,fawcett20,rivera21,fawcett22,calistrorivera23}{}{}. This evidence supports the quasar evolution model proposed by \citet{sanders88}, where the dichotomy between the red and blue quasars arises during an evolutionary phase that connects dust-rich star formation and AGN activity through gas feedback between AGN and its host galaxy. According to the model, within the red quasar phase, the wind/outflow of the central black hole gradually drives away the obscuring dust generated during a period of fast star formation activity (a starburst phase, perhaps driven by merger activity). Eventually, the outflow shuts down SF and reveals the unobscured central black hole, which appears as one of the typical blue quasars \citep[see also][]{hopkins06,hopkins08,farrah12,glikman12}{}{}. \par

Unfortunately, these theories lack direct evidence to validate themselves. \citet{klindt19} and \citet{fawcett20} found that the excess in radio emission of red quasars is mostly seen in compact and radio-faint systems, which often lie around the traditional threshold between radio-quiet and radio-loud sources. \citet{rosario20} used more recent radio data from the LOFAR Two-metre Sky Survey data release 1 \citep[LoTSS DR1;][]{lotssdr1} and found the modelled SF contribution to the total radio emission shows little difference between red quasars and blue quasars, thus concluding that the reddening is likely linked to AGN activities within the system. This argument is further supported by \citet{rosario21} where they used high resolution e-MERLIN data to show that the radio emission in red quasars is more extended on the most compact scales, indicating a greater AGN contribution to the radio emission of red quasars compared to blue quasars. \citet{fawcett22} used data from the \textit{X-shooter} spectrograph and argued that a dusty environment can fully explain most of the colour differences between red and blue quasars, while the radio excess is more likely to be connected with jet interactions in a higher-opacity interstellar medium (ISM)/circumnuclear environment rather than accretion disk activities or outflows. However, all of these studies were unable to separate the contribution from SF and AGN activities due to the shortage of sufficiently high quality radio data.\par

In this work, we adopt the assumptions from the two-component model in \citetalias{macfarlane21} and propose an improved parametric model using a Bayesian approach that can let us study radio emission from the host galaxy SF and AGN jet activity independently within a wider parameter space while obtaining robust results on the possible correlations. Increased quasar samples in the LoTSS DR2 \citep{lotssdr2} and SDSS DR16Q \citep{dr16q} catalogues allow us to further investigate the contribution to the quasar radio emission by any physical processes, including those associated with the quasar colour. As a result, we can finally provide a quantified view of the leading powering mechanism of RQ quasars and the nature of the radio excess in red quasars. \par

This paper is structured as follows: Section~\ref{sec:data} describes the data we used to build a quasar model with multiband measurements; Section~\ref{sec:model} explains the parametric model we proposed to characterise quasar radio emission and the validation of our model; Section~\ref{sec:result_evo} presents our improved result on the dependence of quasar radio emission on optical luminosity and redshift; Section~\ref{sec:colour} shows our result and discussion on the origin of radio excess in red quasars; finally, a summary of our conclusions can be found in Section~\ref{sec:conclusion}. Throughout this work, we assume a $\mathrm{\Lambda CDM}$ cosmology with parameter values published in the WMAP9 result \citep[][]{wmap9}{}{}.

\section{Data}
\label{sec:data}

\subsection{LOFAR Two-metre Sky Survey (LoTSS)}
\label{sec:data_lotss} % used for referring to this section from elsewhere

\begin{figure}     %Insert a figure as soon as possible
    \includegraphics[width=\columnwidth]{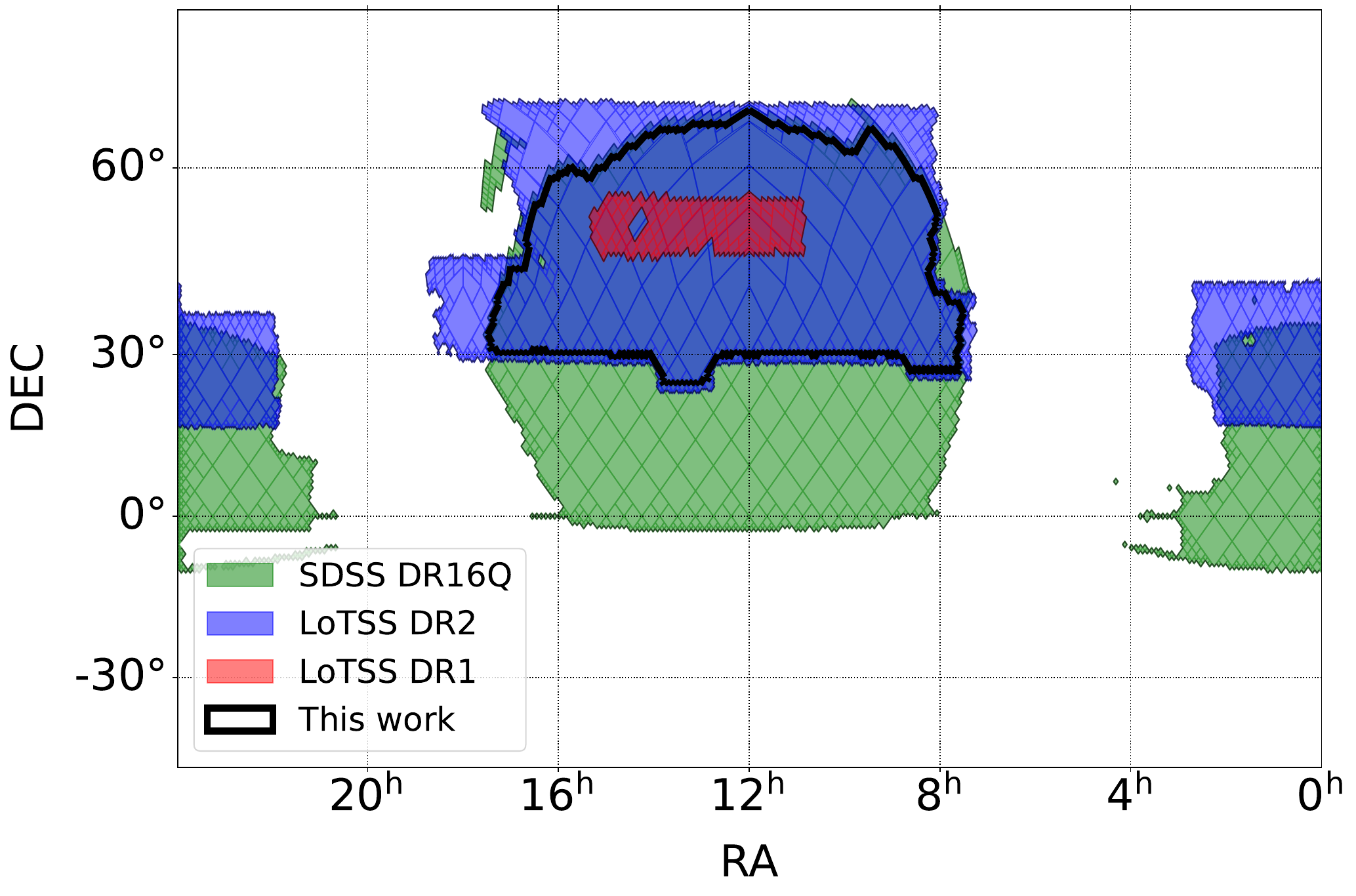}
    \caption{A comparison of the sky coverages from LoTSS DR1 (red), LoTSS DR2 (blue) and SDSS DR16Q (green) survey catalogues. The black line encompasses the sky area studied in this paper, which is the overlap between the LoTSS DR2 and SDSS DR16Q survey areas in the north galactic cap. LoTSS DR2 contains $\sim$4,400,000 radio sources — a ten times increase from the DR1 catalogue, which is ideal for multi-wavelength studies.}
    \label{fig:lotssdr2sky}
\end{figure}

LoTSS \citep[][]{lotss, lotssdr1, lotssdr2} is the LOFAR HBA (high-band antenna) wide-field imaging survey that aims to cover the entire Northern sky in the 120-168 MHz radio band, with a target sensitivity of $\sim100\mu\mathrm{Jy\ beam}^{-1}$ RMS, an angular resolution of $6 \arcsec$ and a positional accuracy of $<0.2 \arcsec$. To date, the most complete catalogue is the LoTSS DR2 catalogue published by \citet{lotssdr2}, spanning over 5,720 $\mathrm{deg}^2$ of sky area and reaching a median sensitivity of $83\mu\mathrm{Jy\ beam}^{-1}$. The LoTSS DR2 survey has $\sim10$ times more coverage compared to the previous data release, DR1 \citep{lotssdr1}, which covers $424~\mathrm{deg}^2$. When compared to the Faint Images of the Radio Sky at Twenty Centimetres survey \citep[FIRST;][]{first}, LoTSS DR2 reaches $\sim10$ times better sensitivity, assuming a spectral slope of $\alpha=0.7$. In total, the LoTSS DR2 catalogue contains $\sim4,400,000$ radio-detected sources. The sky coverage (compared with LoTSS DR1 and the SDSS DR16 quasar catalogues) is shown in Fig.~\ref{fig:lotssdr2sky}. Most of the sky area in LoTSS DR2 is also covered by SDSS DR16 quasar catalogue (described in the section below), providing an ideal combination for multi-wavelength studies. \par
Radio sources in the LoTSS DR2 catalogue were extracted from LoTSS images using the Python Blob Detector and Source Finder \citep[PyBDSF;][]{pybdsf}, identifying sources with peak radio flux densities above the $5\sigma$ limit of the LoTSS DR2 images. While PyBDSF detects regions of radio emission, they are not always correctly grouped into physical sources. To prevent wrong association of the radio sources by the PyBDSF classification, both statistical techniques and extensive visual inspection (using LOFAR Galaxy Zoo) have been used to ensure the radio catalogue represents the actual distribution of the radio sources \citep{hardcastle23}. In addition to the updated radio catalogue, Hardcastle et al. present the cross-matching of LoTSS DR2 sources with optical-infrared counterparts from WISE \citep{wise} and DESI Legacy Imaging \citep{desilegacy} surveys using a method similar to that described in \citet{williams19} and \citet{kondapally21}. \par

\subsection{Sloan Digital Sky Survey (SDSS) quasar sample}
\label{sec:data_sdss}

Optical data for the sample of quasars is drawn from SDSS DR16 quasar catalogue \citep[DR16Q;][]{dr16q}. The DR16Q catalogue is created based on observations from SDSS-I/II/III and IV epochs, reduced using the final eBOSS SDSS reduction pipeline (v5\_13\_0), and spectrally confirmed using the criteria provided in Table 1 of \citet{dr14q}. It also includes previously detected sources from the DR14Q, DR12Q and DR7Q catalogues \citep{sdssdr12q,sdssdr7q}. For quasars with redshift between $0.8<z<2.2$, the quasar classification is done via the decision tree presented in \citet{eboss}; for quasars with redshift $z>2.2$, the classification uses Ly$\alpha$ forest measurements in \citet{myers15} instead. Our quasar optical properties, including \emph{i}-band magnitude and $(g-i)$ colours, show no signs of variation between different selection methods at $z=2.2$. The total number of detected quasars is $>480,000$ and $>239,000$ for the two categories, respectively. Additional optical properties of quasars from the SDSS data, including absolute \emph{i}-band magnitude\footnote{The \emph{i}-band magnitudes presented in the DR16Q catalogue were K-corrected to $z=2$, while in our catalogue they are K-corrected to $z=0$ using the original data and assuming a spectral index of 0.5 \citep{richards06}.} are supplemented by survey data including GALEX, UKIDSS, WISE, FIRST, 2MASS, XMM-Newton and Gaia through cross-matching between catalogues. \par

\subsection{Building a LoTSS-SDSS quasar sample}
\label{sec:data_build}

We aim to build our quasar sample set by extracting LoTSS radio flux density measurements for parent SDSS DR16Q quasar samples; we therefore cross-match the extracted radio data with SDSS optical data based on the quasar sky positions. Before creating our cross-matched catalogue, we applied several extra restrictions to the parent SDSS dataset; these are the same cuts that were applied to the LoTSS DR1 sample in \citetalias{macfarlane21}: \par

\begin{enumerate}
    \item Sources with absolute \emph{i}-band magnitude brighter than -40 were discarded, since they were likely artefacts from the SDSS pipeline. \par 

    \item Sources with redshift $z>4$ were discarded due to the small number statistics and larger possibility of quasar misidentification and/or erroneous redshift measurement. \par

    \item As shown on Fig.~\ref{fig:lotssdr2sky}, only SDSS sources that fell in the 13h field of LoTSS DR2 were included. Sources in the 0h field have a systematically larger uncertainty in radio flux density compared to the 13h field, and are therefore discarded to ensure the conformity of our sample. Other SDSS quasars were excluded if they fall outside the LoTSS coverage, either because of being outside the target field or being in the gaps between the LoTSS mosaics. \par

    \item Finally, another 199 sources in the SDSS DR7Q and DR14Q catalogues were removed; they were selected in the SDSS catalogue only because of their radio emission, and removing them would prevent selection bias toward radio-loud sources outside the SDSS colour-selection region. \par
\end{enumerate}

\begin{figure}     %Insert a figure as soon as possible
    \includegraphics[width=\columnwidth]{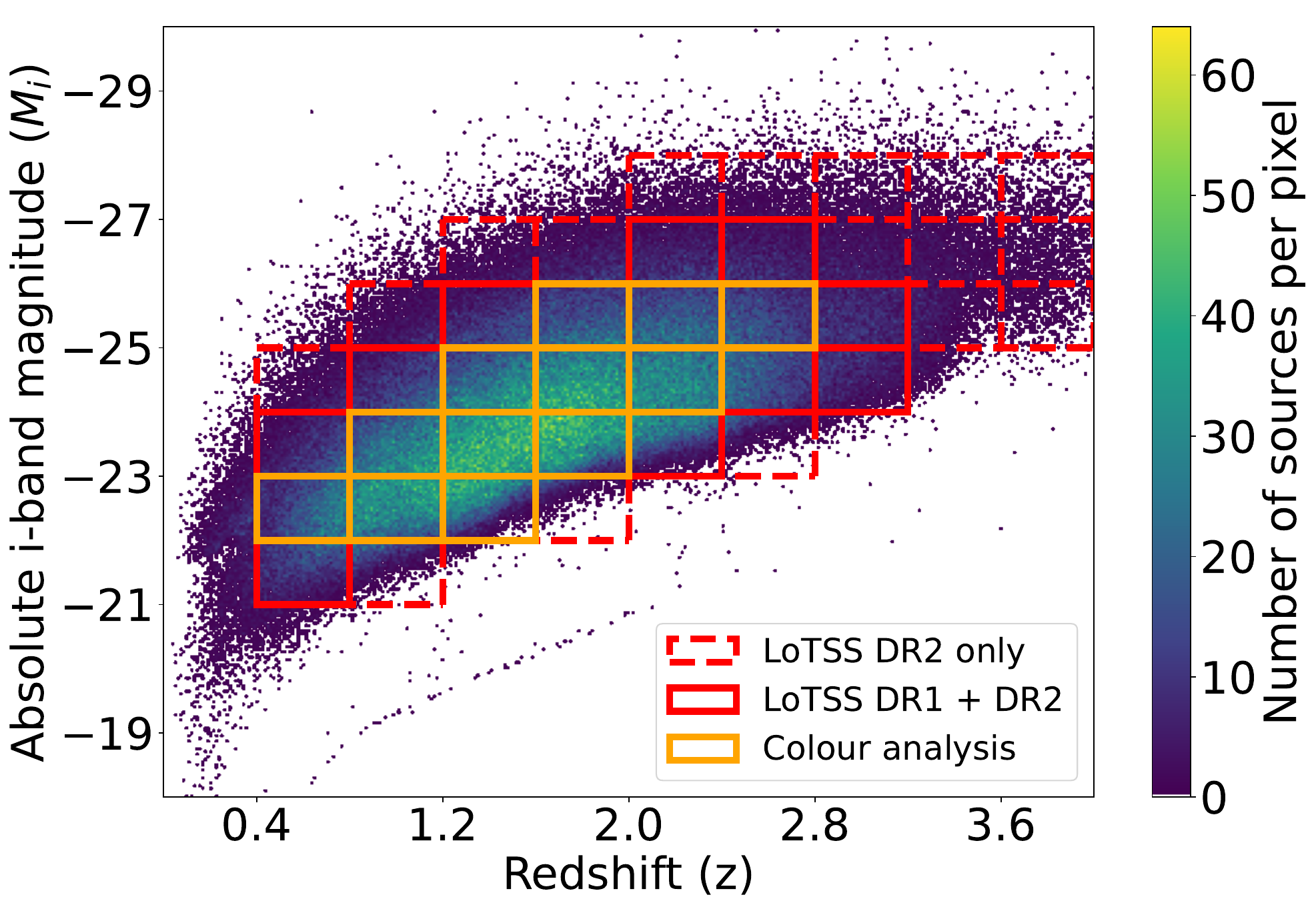}
    \caption{Distribution of sources used in this work in the absolute i-band magnitude ($M_i$)—redshift (z) plane. The samples are later separated into subsets in order to study the dependency of best-fit model parameter on quasar optical properties, using the grid lines in this plot. The grid with solid lines shows the parameter space explored in \citet{macfarlane21} with LoTSS DR1 data, while the dashed grids show the new parameter space explored in this work. Grids with orange lines indicate the parameter space used to investigate the colour dependency in this work, i.e. grid cells with more than 10,000 quasars. Thanks to the new data from LoTSS DR2, we are able to investigate sources with fainter radio emissions and higher redshifts.}
    \label{fig:opticalproperty}
\end{figure}

The final sample consists of 361,123 quasars, with a sample size nearly 10 times larger than the 42,601 quasars used in \citetalias{macfarlane21}. The samples are characterized by their distribution in the \emph{i}-band magnitude ($M_i$)—redshift ($z$) plane, as shown in Fig.~\ref{fig:opticalproperty}. While the sample is naturally biased towards brighter magnitudes at higher redshifts, it still maintains a good dynamical range across the entire parameter space.

\subsection{Quasar properties}
\label{sec:data_properties}

\subsubsection{Radio Flux Densities from LoTSS}
To obtain LoTSS radio flux density measurements for the selected quasars, we adopted different strategies as stated below: \par

\begin{enumerate}
    \item We cross-matched their sky positions in the SDSS catalogue with those in the LoTSS catalogue. This will include quasars from our parent sample with radio flux densities above the LoTSS DR2 $5\sigma$ detection limit. Based on the test result from \citetalias{macfarlane21}, we have set the matching radius to $1.5\arcsec$ for sources above $5\sigma$ detection in LoTSS, in order to balance between the selection rate and random contamination. It is worth noting that the coordinates from the cross-matched optical catalogue of \citet{hardcastle23} were used whenever possible for the LoTSS positions, since these coordinates are more accurate than the flux-weighted radio positions obtained from the PyBDSF code. We have successfully cross-matched 47,902 quasars from this step, all of which are included in our catalogue. \par
    
    \item For quasars with radio flux densities below the LoTSS $5\sigma$ detection limit and which thus do not have a match in the LoTSS DR2 catalogue, we performed a forced photometry on LoTSS mosaics using the method described in \citet{gloudemans21}. This step is motivated by \citet{roseboom14} and \citetalias{macfarlane21} where considerable information has been extracted from sources with detected radio emission dominated by noise. The extracted radio flux density is assumed to be the absolute peak pixel value within the $3\mathrm{px}\times 3\mathrm{px}$ box (where pixels in the LoTSS mosaic are 1.5 arcsec in size, compared to a beam size of $\approx$6 arcsec) surrounding the corresponding position of the source in the SDSS catalogue; the flux density uncertainty is determined by the standard deviation of pixel values in a cutout region of $100\mathrm{px}\times100\mathrm{px}$ surrounding the central pixel. By performing this extraction, we assume the sources undetected by LoTSS are all compact, spatially-unresolved sources; this assumption is reasonable since these radio-quiet sources are likely to be tracing star-formation on galaxy scales, compact radio cores and/or small scale jets -- all of them being relatively small-scaled compared to the $6''$ beam size of LoTSS survey. In fact, only 814 out of 361,123 targets have multiple radio components in the LoTSS DR2 images according to the final optical cross-matched LoTSS DR2 catalogue \citep{hardcastle23}; they are also among the most radio-bright sources in our catalogue, therefore having minimal effect on our results. We have extracted radio flux densities of the remaining 313,221 quasars using this process, bringing the total sample size to 361,123. \par
\end{enumerate}

\subsubsection{Bolometric Luminosities}
\label{sec:data_properties_lbol}
We follow the approach in \citetalias{macfarlane21} and use the absolute $i$-band magnitude ($M_i$) as a proxy for bolometric luminosity of the quasar, and hence the black hole accretion rate. \par

To convert $M_i$ into bolometric luminosity, it can first be converted into the absolute $B$-band magnitude ($M_B$) using the empirical relationship in \citet{richards06}: $M_B-M_i(z=2)=0.66\pm0.31$. The bolometric luminosity is then estimated from the absolute $B$-band magnitude based on the relation in \citet{mclure04}: $M_B=-2.66(\pm0.05)\log\left[L_\mathrm{bol}/\mathrm{W}\right]+79.36(\pm1.98)$. We note that the usage of the absolute $i$-band magnitude as a proxy for absolute magnitude does not take into account possible effects of dust obscuration for the quasars. For most quasars this is expected to be small, but this may be more significant in red quasars. We discuss this further in Section~\ref{sec:colour}. \par

\subsubsection{Colours}
We obtained the quasar $(g-i)$ colours for samples with $z\leq2.5$ from the observed SDSS $g-$ and $i-$band magnitudes. We then used these colours to estimate the dust extinction for quasars across the entire redshift range, and hence the $E(B-V)$ colour excess, which we used to describe the colour-dependent properties of our quasar sample. To be more specific, we compared the colour against the redshift-dependent modal colour of $(g-r)$ and $(r-i)$ bands published in Table 1 of \citet{hopkins04} which therefore accounts for K-correction effects, and used the colour difference between the modal colour and observed $(g-i)$ colour to obtain the reddening (relative to the average quasar) in the $(g-i)$ bands. Following the work in \citet{glikman22} using WISE-2MASS selected quasars, we used an SMC\footnote{Small Magellanic Cloud}-like extinction law \citep[][]{pei92} to convert the reddening into the relative colour excess $\Delta E(B-V)$ of the quasar samples. Note that while these $E(B-V)$ values (resembling the difference from the average colour) will scatter around zero with apparently unphysical negative values indicating less reddening than the average quasar, our interest is in the tail towards the red end, where the choice to use $\Delta E(B-V)$ allows us to more easily compare across different redshift ranges.\par

\section{Parametric model of observed radio flux density distribution}
\label{sec:model}

To construct a statistical model that translates the observed \emph{radio flux density} distribution to a parametric likelihood, we adopt the framework presented in \citet{roseboom14} to allow the statistical likelihood of each individual source to be stacked together and modelled as an ensemble. This method makes full use of every source available and allows the overall probability to be constrained consistently. We will briefly summarize the \citet{roseboom14} approach below and then show how it can be adapted into our work. \par

The radio data consists of a set of sources with sky positions $\boldsymbol{x}$ and observed flux densities $\boldsymbol{d}_x$. Instead of using a traditional approach to estimate the model parameters \citepalias[e.g.][]{macfarlane21}, we instead use a bottom-up approach by assuming the underlying flux densities to be $\boldsymbol{s}_x$, in which case the pixel intensity can be modelled as:

\begin{equation}
    \boldsymbol{d}_x=\boldsymbol{s}_x+\boldsymbol{\delta}_x,
    \label{eq:3.2.1}
\end{equation} 

\noindent where $\boldsymbol{\delta}_x$ is a series of random statistical errors drawn from a Gaussian distribution $N(\mu=0, \sigma_s)$, and $\sigma_s$ represents the different uncertainties on the flux density measurements for each sample. If a certain model $M$ can be constructed to predict the probability $P(s)$ of a source having \textit{underlying} flux density $s$, then the probability of observing a pixel intensity $d$ for that source is:

\begin{equation}
    P(d|M,\sigma_s)=\int_0^\infty P(s)\cdot\frac{1}{\sigma_s\sqrt{2\pi}}\exp\left(\frac{-(s-d)^2}{2\sigma_s^2}\right)ds,
    \label{eq:3.2.2}
\end{equation}

\noindent by convolving equation~(\ref{eq:3.2.1}) over the entire parameter space for $s$, folding in the Gaussian noise distribution. The stacked probability of a population with observed flux density distribution $\boldsymbol{d}_x$ and RMS flux density error $\boldsymbol{\sigma}_{s,x}$ can then be expressed as:

\begin{equation}
    P(\boldsymbol{d}_x|M,\boldsymbol{\sigma}_{s,x})=\prod_x P(d|M,\sigma_s).
    \label{eq:3.2.3}
\end{equation}

Applying Bayes theorem, we can derive the likelihood of the model M based on the current distribution $\boldsymbol{d}_x$:

\begin{equation}
    P(M|\boldsymbol{d}_x,\boldsymbol{\sigma}_{s,x})\propto P(M)P(\boldsymbol{d}_x|M,\boldsymbol{\sigma}_{s,x}),
    \label{eq:3.2.4}
\end{equation}

\noindent thus, the best-fitting parameters of model M can be found by maximizing the stacked likelihood in equation~(\ref{eq:3.2.3}) under any given prior to the model ($\rho(M)$). \par

\subsection{Two-component model for radio emission}
\label{sec:model_twocomp}
To construct the parametric expression of the probability of a source having true radio flux density $s$ ($P(s)$), we base our approach on the two-component model proposed in \citetalias{macfarlane21}. Each quasar in the sample is assumed to have contributions to the radio flux density from both star-forming activity and AGN (jet) activity, while the total sample set is binned into grid cells on the $M_i-z$ plane, each of which is modelled separately (see Figure~\ref{fig:opticalproperty}). Here, we will first introduce the two model components individually, before explaining how we will combine them into a parametric expression of $P(s)$.

\subsubsection{SF Component}

The host galaxies for radiative-mode AGNs are known to be massive and star-forming \citep[e.g.][]{kauffmann03,best05}. \citetalias{macfarlane21} modelled the radio emission arising from the host galaxy star-forming activities within each grid cell with a log-Gaussian probability distribution centred at a certain 150 MHz radio luminosity, $\log(L_\mu/[\mathrm{W~Hz}^{-1}])$ (tracing typical SFR $\Psi$ within this population), and a scatter, $\sigma_\mu/\mathrm{dex}$. The conversion between SFR ($\Psi$) and radio luminosity ($L_\mu$) is provided by \citet{smith21}, being calibrated with LoTSS data and a \citet{chabrier03} initial mass function:

\begin{equation} 
    \log\left(\frac{\Psi}{M_\odot\mathrm{yr}^{-1}}\right)=0.96\left[\log\left(\frac{L_\mu}{\mathrm{W~Hz}^{-1}}\right)-22.181\right],
    \label{eq:3.2.6}
\end{equation}

\noindent with scatters within 0.3 dex in our typical SF galaxy radio luminosity range $L_\mu\sim10^{23-24}\mathrm{W~Hz}^{-1}$. The scatter for SFR is propagated from the scatter in luminosity: $\sigma_\Psi=0.96\sigma_\mu$. Note that while there is some evidence for a mass-dependent SFR-radio luminosity relation \citep[e.g.][]{algera20,delvecchio21,smith21}{}{}, we use the mass-independent form of the correlation in \citet{smith21} since we do not have robust measurements for the host galaxy stellar mass. Given the tight correlation between SFR and stellar mass at a fixed redshift \citep[e.g.][and references therein]{speagle14} and that the estimated SFRs for the quasar sample are relatively high (see Section~\ref{sec:result_evo}), we can infer relatively high stellar masses ($>10^{10} M_\odot$) within a given bin. Therefore, the weak mass dependency in the \citet{smith21} relation is unlikely to have a significant impact on the result when compared to the intrinsic scatter in the $L_{150}$-SFR correlation.\par

Since the physical properties are expected to be similar for sources sharing a grid cell on the $M_i-z$ plane, we expect a tight correlation between radio luminosity and star formation rate if either SFR correlates with gas accretion rate, or if host galaxies lie on the SF main sequence. This provides a broad justification for the log-Gaussian model we use here, and results from \citetalias{macfarlane21} further support the feasibility of our model. Note that values of $L_\mu$ and $\sigma_\mu$ may vary between different grid squares: the normalisation of the star-forming main sequence has a strong dependence on redshift, due to more gas being available at the peak of the cosmic star formation rate density \citep[e.g.][]{madau14}; many studies have also found variance between typical host galaxy SFRs with AGN luminosities \citep[e.g.][]{shao10,bonfield11,rosario12,dong16}{}{}. \par

The probability distribution function (PDF) for the star formation component thus becomes (where L is the radio luminosity of the source):

\begin{equation} 
    P_\mathrm{SF}(L)dL=\left[\frac{1}{\sigma_\mu\sqrt{2\pi}}\exp\left(\frac{\log L-\log L_\mu}{2\sigma_\mu^2}\right)\right]\frac{dL}{L}.
    \label{eq:3.2.7}
\end{equation}

\par

\subsubsection{AGN Component}
Since the radio luminosity of radio-loud AGN is dominated by their jet component, we can extrapolate the luminosity function of radio-loud AGNs to lower luminosities to reflect the distribution of jet luminosities for both radio-loud sources and radio-quiet sources, assuming they share the same jet mechanism, only with varying power efficiency. Note that while there are several other proposed scenarios for AGN-related radio emission in RQ AGNs, we only consider the AGN jets here for simplicity. Impacts from other mechanisms are negligible in the radio-bright end of our model, but might affect the radio-faint end; we will further discuss such impacts in Section~\ref{sec:colour}. \par

\citetalias{macfarlane21} used a single power-law distribution to model the emission from radio-loud AGNs (and thus the jet component emission), which results in a probability distribution function such that:

\begin{equation} 
    P_\mathrm{jet}(L)dL=\phi L^{-(\gamma-1)}\cdot\frac{dL}{L}\quad\quad(L>L_\mathrm{jet}^\mathrm{min}),
    \label{eq:3.2.8}
\end{equation}

\noindent where $\gamma$ is the power-law slope, $\phi$ is the normalisation parameter which can be translated into the radio-loud fraction $f$ defined in \citetalias{macfarlane21} (see the discussion in Section~\ref{subsec:model_buildup}), and $L_\mathrm{jet}^\mathrm{min}$ is the lower limit of the jet luminosity, required for normalisation (see below). \par

While the luminosity function for radio-loud AGNs is often modelled as a broken power-law distribution \citep[e.g.][]{dunlop90}, this luminosity function traces the entire population of radio AGNs; therefore, quasars with different optical luminosities may occupy different parts of the function. As a result, if we are only sampling a part of the entire population in a given grid cell, the distribution may not follow the same pattern as the integrated population. \citetalias{macfarlane21} found that the lack of high-luminosity sources in a number of grid cells caused strong degeneracies between parameters, and the slope above the break luminosity could not be constrained. Here we adopt the \citetalias{macfarlane21} conclusion and use a single power-law function to model the jet component instead. We demonstrate later that this provides a very good fit to the data.\par

A practical issue with a power-law probability distribution function is that it grows monotonically when moving towards the faint end. To tackle this problem, \citetalias{macfarlane21} set a lower luminosity limit ($L_\mathrm{jet}^\mathrm{min}$ in Equation~\ref{eq:3.2.8}) for all quasars (below which the probability dropped to zero) such that the integral of the jet probability distribution comes to unity. They found that this lower luminosity limit is typically $10^{19}\sim10^{20}\ \mathrm{W~Hz^{-1}}$, which is in line with the expectations from previous models \citep[e.g.][]{mauch07, cattaneo09, sabater19}{}{}. \par

\subsubsection{Total Simulated Radio Emission}
\label{subsec:model_buildup}
While \citetalias{macfarlane21} randomly drew an $L_\mathrm{SF}$ and $L_\mathrm{jet}$ from the corresponding PDF and summed them in Monte Carlo simulations to get the overall luminosity distribution, in this work we need to derive a PDF for the \emph{individual} quasar luminosity that we can convert to the probability function $P(s)$ in Equation~\ref{eq:3.2.2}. We take a different approach of re-scaling and summing the SF and jet component PDFs (rather than summing luminosities drawn from the PDFs), as outlined below. \par

Under the two-component assumption, when the jet is very weak, the quasar luminosity will be dominated by SF emission; thus its distribution will follow the distribution function of the SF component. However, the growth of the jet distribution function at the faint end would instead let the jet component dominate at low flux densities if the two PDFs are simply summed, giving a clearly incorrect combination. We hereby make a simplifying assumption of letting $L_\mathrm{jet}^\mathrm{min}=L_\mu$ in Equation~\ref{eq:3.2.8}; we then rescale and add the two probability distributions (Equation~\ref{eq:3.2.7} and~\ref{eq:3.2.8}) together to form the probability density distribution for the \textit{total} underlying radio luminosity $P(L)$. \par

Under this assumption, when the jet luminosity falls below $L_\mu$, $P(L)$ is proportional to the PDF of the SF component: $P(L)\propto P_\textrm{SF}(L)$, since the SF component already dominates the entire luminosity distribution. On the other hand, when $L_\textrm{jet}\gg L_\mu$ the jet component becomes dominant in the distribution function and the overall PDF becomes the scaled version of $P_\textrm{jet}(L)$. \par

To achieve this in practice, for any given jet distribution, we let $\eta$ be the fraction of the sources that have $L_\textrm{jet}>L_\mu$. The value of the AGN-dominated source fraction $\eta$ is given by: 

\begin{equation}
  \label{eqn-eta}
    \begin{split}
        \eta = \int_{L_\mu}^{\infty} P_\textrm{jet}(L)dL 
        = \int_{L_\mu}^{\infty} \phi L^{-\gamma}dL 
        = \phi\cdot\frac{L_\mu^{1-\gamma}}{\gamma-1}.
    \end{split}
\end{equation}

\noindent Since the jet component dominates the luminosity distribution of these sources, we use $P(L)=\eta\cdot P_\textrm{jet}^*(L)$ to describe their luminosity function, where $P_\textrm{jet}^*(L)=P_\textrm{jet}(L)$ for $L_\textrm{jet}>L_\mu$ and $P_\textrm{jet}^*(L)=0$ for $L_\textrm{jet}<L_\mu$. For the remaining $(1-\eta)$ fraction of the sources, the total luminosity is dominated by the SF contribution. Therefore, for these sources, we have $P(L)=(1-\eta)P_\textrm{SF}$. We thus define the luminosity function (or overall PDF) in our model as:

\begin{equation} 
    P(L)dL = [(1-\eta)P_\mathrm{SF}(L)+\eta P_\mathrm{jet}^*(L)]dL.
    \label{eq:3.2.9}
\end{equation}

\noindent For $L\leq L_\mu$ we therefore have:

\begin{equation}
    P(L)dL=\frac{1}{\sigma_\mu\sqrt{2\pi}}\exp\left(\frac{\log L-\log L_\mu}{2\sigma_\mu^2}\right)\frac{dL}{L};
    \label{eq:3.2.10}
\end{equation}

\noindent and for $L>L_\mu$ we have:

\begin{equation}
    P(L)dL = \left[\frac{1-\eta}{\sigma_\mu\sqrt{2\pi}}\exp\left(\frac{\log L-\log L_\mu}{2\sigma_\mu^2}\right)+\eta\phi L^{-(\gamma-1)}\right]\frac{dL}{L}.
    \label{eq:3.2.11}
\end{equation}

It is also worth noting the relationships between the jet normalisation parameter $\phi$ or the AGN-dominated fraction $\eta$ used in this work and the similar parameter $f$ defined by \citetalias{macfarlane21}. The parameter $f$ in \citetalias{macfarlane21} is defined as the fraction of sources with jet luminosity brighter than $L_f$, where $\log(L_f/\mathrm{W~Hz^{-1}})=26$. Therefore, we have $f=\int_{L_f}^{\infty} \phi L^{-\gamma}dL$. Our definition of $\eta$, on the other hand, gives $\eta=\int_{L_\mu}^{\infty} \phi L^{-\gamma}dL$. We can thus convert between $\eta$ and $f$ using: 

\begin{equation}
    \frac{f}{\eta}=\frac{\int_{L_f}^{\infty} \phi L^{-\gamma}dL}{\int_{L_\mu}^{\infty} \phi L^{-\gamma}dL}=\left(\frac{L_f}{L_\mu}\right)^{1-\gamma}.
\end{equation}

\noindent Combining this scaling relation with Equation~(\ref{eqn-eta}) gives the following conversion between the jet normalisation parameter $\phi$ used in this work and the parameter $f$ in \citetalias{macfarlane21}:

\begin{equation}
    \phi=(\gamma-1)L_f^{\gamma-1}\cdot f.
    \label{eq:convert_logphi}
\end{equation}

Having established the PDF for radio luminosities, $P(L)$, the probability distribution of \emph{radio flux densities} $P(s)$ in Equation~(\ref{eq:3.2.2}) (assuming fixed redshift within each grid) becomes:

\begin{equation} 
    P(s)=\Phi(4\pi D(z)^2k(z) s),
    \label{eq:3.2.13}
\end{equation}

\noindent where $D(z)$ is the luminosity distance at redshift $z$, and $k(z)$ is the K-correction\footnote{We adopt a radio spectral slope of $\alpha=0.73$ (assuming $L\propto\nu^{-\alpha}$), as in \citet{rivera17}. Thus, we have $k(z)=(1+z)^{\alpha-1}$.} applied at $z$. Note that here $P(s)$ is the probability of a \emph{particular} quasar having an \emph{underlying} radio flux density $s$. \par

We refer the reader to Table~\ref{tab:params} for a brief summary of the model parameters presented above.

\begin{figure*}     %Insert a figure as soon as possible
    \includegraphics[width=\textwidth]{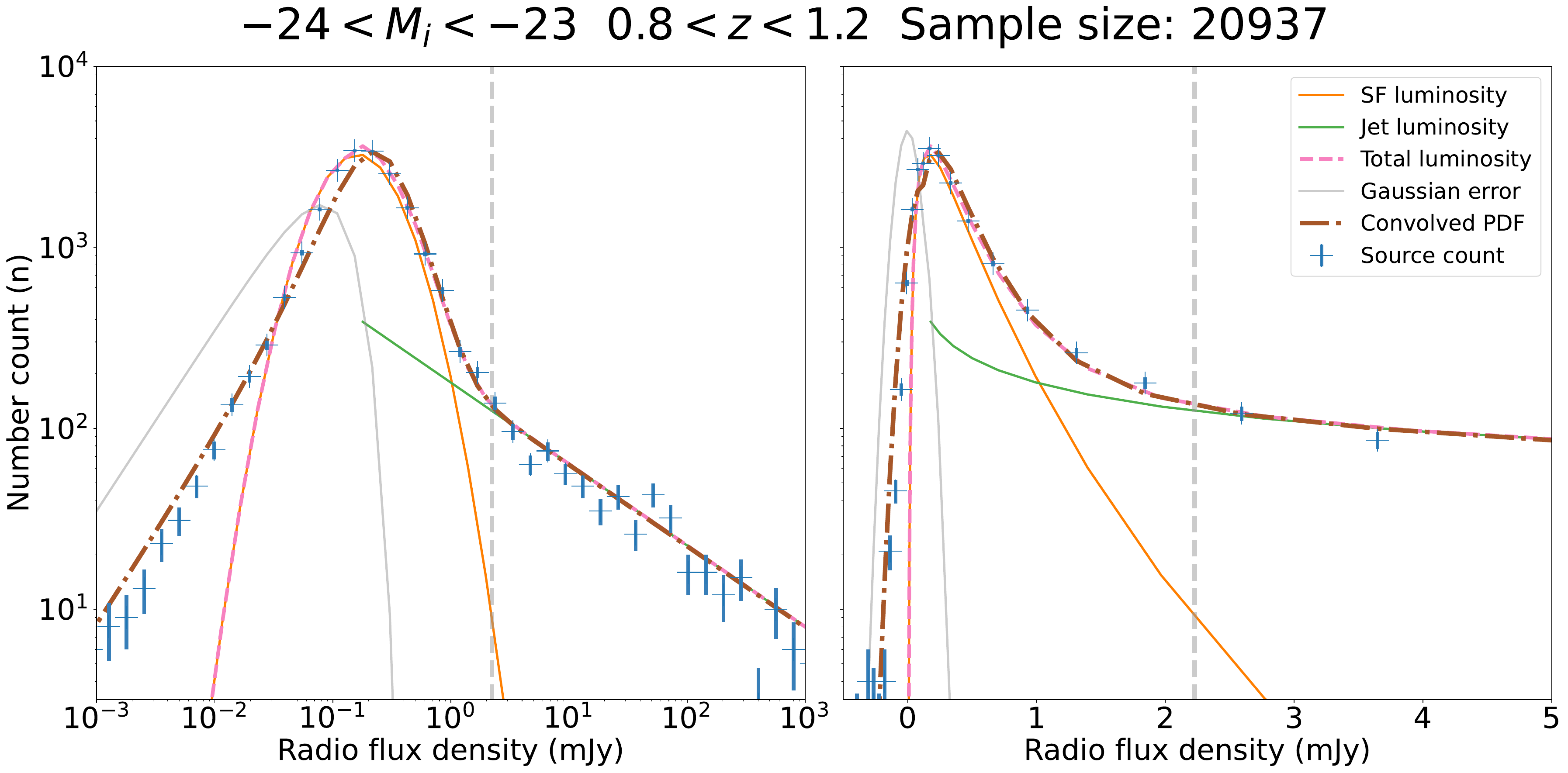}
    \caption{Separate and combined probability distribution functions (PDF) of radio flux densities ($\nu$), illustrating different components in our model. The left panel shows the flux density distribution on a logarithmic scale which offers the clearest representation of the model, while the right panel shows the flux density distribution on a linear scale, which is important to show the negative values (due to noise) and is the form used in our fitting algorithm. In each panel, the orange line and green line represent the SF component from the host galaxy activity (log Gaussian) and the jet component from the AGN activity (single power-law) respectively. The pink dotted line represents the combined PDF of the two-component radio flux density distribution model. The vertical grey dashed line indicates the lower radio luminosity limit that we used to fit $\gamma$ value in this $M_i$-$z$ grid (see Section~\ref{sec:model_fitting} for details). The grey solid line compares the Gaussian error introduced by flux measurement (using average flux error in the extraction from DR2 mosaics) with the modelled distribution and the actual distribution. Note that the actual probability density of our Gaussian flux density error peaks close to zero in linear units of flux density, with the difference between the two panels being that the bin size in the left panel (with logarithmic bins) becomes smaller towards the radio-faint end, hence the number count within each bin will drop with the bin sizes. The red dashed-dotted line shows the PDF after convolving with the observational error. We have drawn sources from a grid cell in our catalogue ($-24<M_i<-23$, $0.8<z<1.2$) and binned them in radio flux density to get the observed number count within each bin (shown in blue crosses, with the uncertainty shown in thick blue lines along the y-axis). The observed distribution agrees well with the proposed model.}
    \label{fig:pdf}
\end{figure*}

\begin{table}
    \caption{A list of our model parameters and their definitions. All luminosities refer to the extracted radio luminosities from LoTSS catalogue.}
    \label{tab:params}
    \begin{tabular}{cc}
        \hline
        $L_\mu/[\mathrm{W~Hz}^{-1}]$ & Average SF luminosity of the population \\
        $\sigma_\mu/\mathrm{dex}$ & Gaussian scatter in the SF luminosity distribution \\
        $\gamma$ & Jet power-law slope as in $P_\mathrm{jet}(L)=\phi L^{-\gamma}$ \\
        $\phi$ & Jet normalisation factor as in $P_\mathrm{jet}(L)=\phi L^{-\gamma}$ \\
        $\eta$ & Fraction of sources with $L_\textrm{jet}>L_\mu$ (AGN-dominated) \\
        \hline
    \end{tabular}
\end{table}

\subsection{Fitting with a Parametric Approach}
\label{sec:model_fitting}
With the expression of $P(s)$, we can finally calculate the stacked likelihood of a certain parameter set $\{L_\mu, \sigma_\mu, \gamma, \phi\}$ under a radio flux distribution $\boldsymbol{d}_x$ and an RMS error $\sigma_s$, using Equations~(\ref{eq:3.2.2}) to~(\ref{eq:3.2.4}). We then use a Monte Carlo Markov Chain (MCMC)-based algorithm \citep[\emph{emcee};][]{emcee} to determine the best-fit parameters based on the marginalised median of the probability density distribution defined in Equation~(\ref{eq:3.2.4}).\par

Figure~\ref{fig:pdf} gives an overview of our parametric model described above, for one well-populated grid cell. The left panel shows the flux density distribution in log space, which explicitly shows the two-component model discussed in Section~\ref{sec:model_twocomp}; the right panel shows the model in linear space — reaching negative values — to better show the effects of the noise. The orange line and green line represent the radio flux density PDF of the log-Gaussian SF component (Equation~\ref{eq:3.2.6}) and the single power-law jet component (Equation~\ref{eq:3.2.8}) respectively. The pink dashed line represents the combined PDF of the two-component model (Equation~\ref{eq:3.2.9}), while the red dashed-dotted line shows the final PDF after convolving with the observational error (Equation~\ref{eq:3.2.13}). We have drawn from the quasar parent sample a grid cell containing 20,937 sources with optical magnitude $-24<M_i<-23$ and redshift $0.8<z<1.2$, and binned them in radio flux density (shown in blue crosses); the best-fit PDF agrees well with the distribution of the observed sources. There is a small mismatch at $L$ just below $L_\mu$ (the blue data points above the brown dashed-dotted line), perhaps caused by our method of combining the PDFs assuming no jet contribution here, while there might still be a weak amount from jet just below $L_\mu$, leading to an underestimation in the constructed PDF. Other well-populated grid cells share a similar pattern, despite being located at different places in the parameter space.\par

To test the limits of viability and robustness of our model within the parameter space explored, we have created mock catalogues with different numbers of sources (ranging from 250 to 10,000) sampled from the luminosity function in Equation~\ref{eq:3.2.11} with different input parameters, and fitted our model with the mock catalogues. We then compare the fitted values of the model parameters against the input to test whether our fitting approach can retrieve the actual parameter values. Details on the validation results using mock data can be found in Appendix~\ref{sec:mocktest}. 

The test results show that the lower limit of $L_\mu$ to which our approach can probe depends on the number of quasars in the bin - having more quasars allows us to probe fainter $L_\mu$. The lower limit of $L_\mu$ corresponds to a flux density around 5 times the stacked noise level. Assuming an average noise level of LoTSS and a typical SFR of the SDSS quasars, our approach accurately retrieves input parameter values for grid cells containing at least 1,000 sources. To comply with the test results, we consider only grid cells with more than 1,000 sources (10,000 sources for analysis regarding other physical parameters, since we subdivide our samples into 10 bins) in the following analysis; this is indicated by the colour-coding of grid squares in Figure~\ref{fig:opticalproperty}.  \par

The mock test results indicate a degeneracy between $\sigma_\mu$ and $\gamma$ in our fitting; this is most likely due to the relative scarcity of sources in the radio-loud regime, which makes the best-fit result of $\gamma$ depend more on the sample distribution in the region where both the SF and jet make a significant contribution to radio emission. On the other hand, assuming a single power-law distribution in the radio-loud quasars provides us a simple way to estimate $\gamma$ without running MCMC fits. As the host galaxy SF contribution becomes insignificant in the radio-bright tail of the distribution, the power-law slope of the distribution at high radio luminosities provides a good estimation for $\gamma$ in our model. We can then use the fitted values of different power-law slopes across different bins as strong informed priors to the $\gamma$ values in our model, and thus resolve the degeneracy between $\sigma_\mu$ and $\gamma$ in the radio-bright end. \par

\begin{figure}
	\includegraphics[width=\columnwidth]{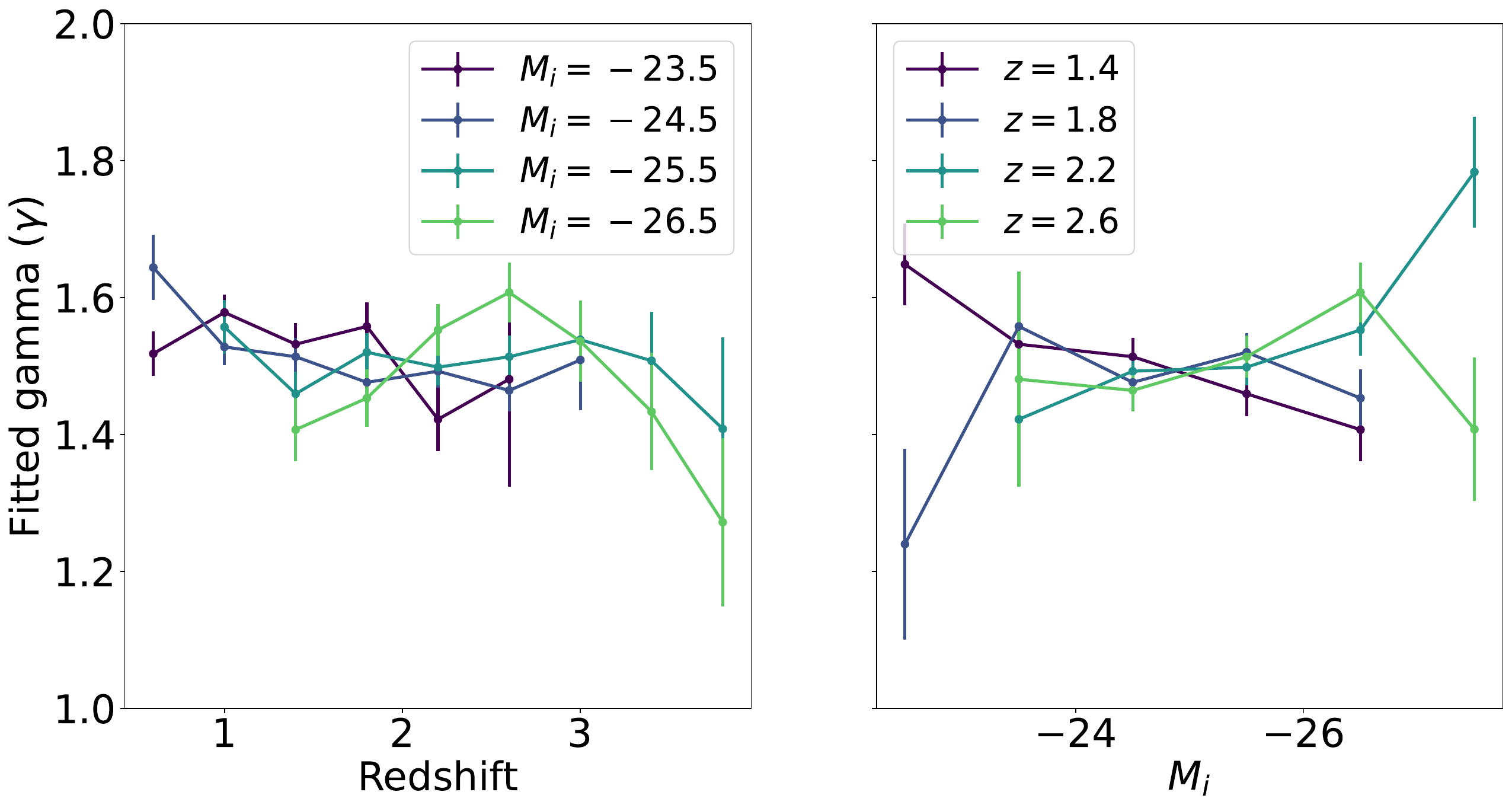}
    \caption{The $\gamma$ value inferred from the power-law slope at the bright end of the radio luminosity distribution, plotted for the most populated grids. The values of $\gamma$ show little change with redshift (left panel) or absolute $i$-band magnitude (right panel), agreeing with the conclusion in \citetalias{macfarlane21}. Therefore, we fixed the value of $\gamma$ to $\gamma \approx 1.5$ when fitting the general sample (not binned with colour), using a tight Gaussian prior.}
    \label{fig:gamma_fit_slice}
\end{figure}

The results of the analysis on the $\gamma$ prior are provided in Appendix~\ref{sec:priortest}, where Figure~\ref{fig:gamma_fit_full} shows the radio-bright end of the quasar radio luminosity function for all grid bins investigated in this work (blue line). The lower luminosity limits for analysis of the radio-bright luminosity functions are selected using the following criteria: (i) not lower than the $10\sigma$ radio flux density limit and (ii) at least 1 dex above the estimated $L_\mu$ to ensure negligible SF contribution, so that the distributions only show a single power-law feature in the $\log n-\log L$ space (values listed in Table~\ref{tab:result}). We used a single power-law model ($n(L)\propto L^{-\gamma}$) to fit the radio-bright distributions, as shown by the red dotted lines in each grid cell. The single power-law model provides a good fit across the entire parameter space, which further justifies our model for the jet components, and therefore supports the ubiquity of jets in our quasar samples. The fitted value of $\gamma$ is shown in Figure~\ref{fig:gamma_fit_full} for every grid cell, while Figure~\ref{fig:gamma_fit_slice} gives an overview of the $\gamma$ values across some of the most populated grids. These values show little dependence on the redshift or optical magnitude, which agrees with the conclusion in \citetalias{macfarlane21} using the chi-square fitting.

As a result, within each grid, the informed prior of $\gamma$ is defined as a Gaussian distribution: \par

\begin{equation} 
    \ln{\rho(M})=-\frac{(\gamma-\gamma_0)^2}{2\sigma_\gamma^2},
    \label{eq:3.2.14}
\end{equation}

\noindent where $\gamma_0$ is fixed at 1.5 at all grid bins due to little evolution with redshift or optical magnitude, and $\sigma_\gamma$ is set to 0.05 to give $\gamma$ a tight range. \par

For the rest of the parameters, we have adopted simple box priors ($19<\log L_\mu<30$, $0.05<\sigma_\mu<0.5$, $-5<\log f<-0.25$) to maximise the information obtained from the MCMC fitting while ruling out the unrealistic results. The final expression for the likelihood of the model is therefore given by Equation~\ref{eq:3.2.4} assuming $P(M)=\rho(M)$.

\section{Dependencies with $M_{\lowercase{i}}$ and \textit{\lowercase{z}}}
\label{sec:result_evo}
In this section, we present the results related to the best-fit parameters from our proposed model. To better compare our results with the previous ones from \citetalias{macfarlane21}, we use the scaling relation in equation~\ref{eq:convert_logphi} to convert the jet power normalisation parameter $\phi$ obtained in this work to the parameter $f$ presented in \citetalias{macfarlane21}. In Section~\ref{sec:model_twocomp} we also proposed a more physically motivated parameter - the jet probability scaling factor ($\eta$) - defined as the fraction of quasars with radio emission brighter than the average host galaxy contribution, on which our analysis will focus mainly.\par

\begin{figure*}
	\includegraphics[width=\textwidth]{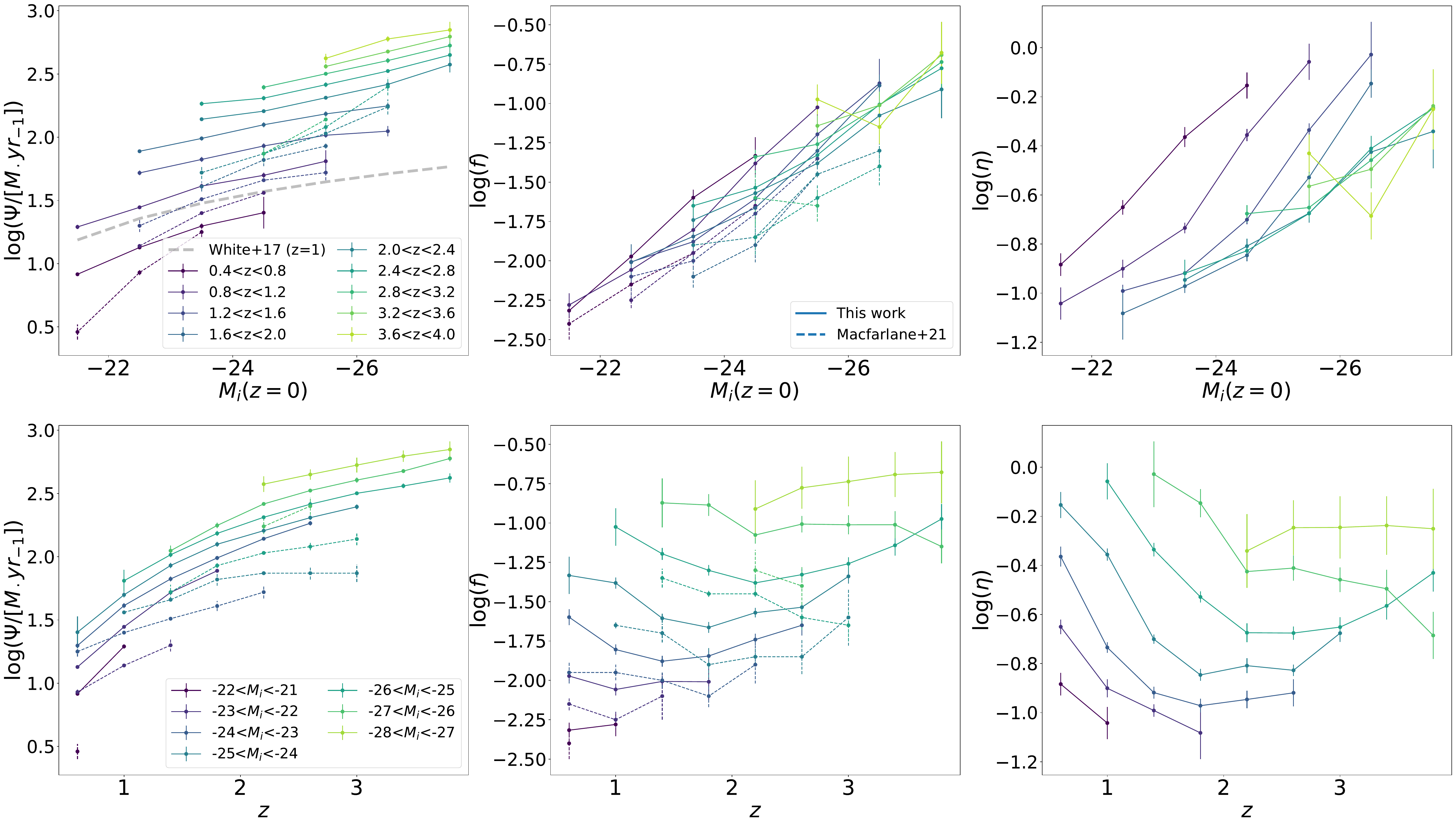}
    \caption{The best-fit values of the mean host galaxy SFR ($\Psi$, as determined from $L_\mu$ using Eq.~\ref{eq:3.2.6}), jet power normalisation from \citet{macfarlane21} ($f$), jet probability scaling factor from this work ($\eta$), and their variation with absolute $i$-band magnitude ($M_i$, top row) and redshift ($z$, bottom row). The solid lines show the result of this work, and we compare our result with previous results in \citetalias{macfarlane21} (dashed lines). In line with \citetalias{macfarlane21}, we observe a significant dependence of both SFR ($\Psi$) and jet activity ($f$ or $\eta$) with $M_i$ (a proxy for bolometric luminosity and accretion rate), while only the SFR evolves strongly as a function of redshift for fixed $M_i$ (see the text for further discussion). Our SFR-$M_i$ correlation at $0.8<z<1.2$ also shows good agreement with the best-fit relationship in \citet{white17} where they studied the radio emission for 70 radio-quiet quasars at $z\sim1$ (grey line). }
    \label{fig:result_full}
\end{figure*}

Figure~\ref{fig:result_full} shows the variation in our fitted parameters across different subsamples binned by their absolute $i$-band magnitudes and redshifts, while we present the detailed values of best-fit parameters in Table~\ref{tab:result}. Compared to previous results in \citetalias{macfarlane21}, our method extended the investigated redshift range to $z\approx3.8$ and the optical luminosity range up to a magnitude deeper, thanks to the wider sky area coverage of LoTSS DR2 and the improved parametric fitting method, which requires fewer samples for a good fit. The fitted values for the star formation rate of the host galaxies ($\Psi$) exhibit a much higher precision when compared to the similar results using chi-square-based approach in \citetalias{macfarlane21} - another major improvement made possible by our Bayesian fitting method. However, the values of $\Psi$ in our work deviate significantly more from the best-fit values in \citetalias{macfarlane21} than the errors, in both faint-$M_i$ and high-$z$ sources. This is believed to be largely due to the differences between our model and the \citetalias{macfarlane21} model. Firstly, we have calculated a higher prior for $\gamma$ ($\gamma=1.5$) compared to the \citetalias{macfarlane21} model ($\gamma=1.4$) due to the difference in our fitting strategy (see \citetalias{macfarlane21} for details), which leads to an increased fraction of jet-dominated sources in the radio-intermediate regime. As a result, the scatters of the SF component ($\sigma$) that we derive is notably smaller compared to \citetalias{macfarlane21} since fewer SF-dominated sources are required in the radio-intermediate quasar population, which would therefore lead to higher fitted values of peak SFR ($\Psi$) to keep a consistent jet normalisation in the radio-loud regime.\footnote{We have tried further comparisons by fixing the $\gamma$ and $\sigma_\mu$ to the values adopted in \citetalias{macfarlane21} and using the parameter space where this work and \citetalias{macfarlane21} overlap, however we did not get a converged fitting result due to the aforementioned fundamental differences in the model definitions. } Secondly, the way that we combine the PDFs for the SF and AGN components is subtly different, which predominantly influences sources around $L_\mu$; where $L_\mu$ corresponds to a flux density below the noise limit, this can impact the results. Thirdly, we convolved our theoretical PDF with different radio flux uncertainties of each individual sources (see Equation~\ref{eq:3.2.2}), while the \citetalias{macfarlane21} model used the average radio flux uncertainty for the full sample; although our method maps the actual physical scenario more accurately, it can also lead to deviations from the \citetalias{macfarlane21} result. However, whilst all of these factors affect our comparison with the of \citetalias{macfarlane21} values, variations between quasar sub-populations within our model are very robustly determined.
\par

The absolute $i$-band magnitude ($M_i$) is a good tracer of the bolometric luminosity, and is hence associated with the BH accretion rate of a quasar (see Section \ref{sec:data_properties_lbol}). Note that although there is a degeneracy between BH mass and BH accretion rate in their correlations with the bolometric luminosity, in this paper we will focus on the BH accretion rate side and leave the investigation of possible correlations with BH mass for future work. In this study, we have observed a positive correlation with $M_i$/$L_\textrm{bol}$ in both host galaxy SFR and AGN activity level (characterised by $\eta$). By assuming $\textrm{SFR}\propto L_\textrm{bol}^\alpha$ under fixed redshifts, we have $\alpha=0.26\pm0.02$, which is in agreement with the value found in \citet{bonfield11} where they used far-IR \textit{Herschel} measurements instead as a SF tracer. We have also compared our result with the $L_\mathrm{SF,1.5GHz}$-$M_i$ correlation presented in \citet{white17}, where they obtained 1.5 GHz radio measurements of 70 radio-quiet quasars at $z\sim1$ using the Karl G. Jansky Very Large Array (JVLA). Their best-fit relationship, with $L_\textrm{SF,1.5GHz}$ converted to a 150 MHz radio SFR using the radio spectral slope defined previously,  is plotted as the grey dashed line in Figure~\ref{fig:result_full}, and is in agreement with our $\textrm{SFR}$-$L_\textrm{bol}$ correlation at $0.8<z<1.2$ (albeit under a different assumed best-fit function). \par

Our results also agree with the results in \citetalias{macfarlane21} when comparing the fitted values of $\log f$. Adopting their definition of $f\propto L_\textrm{bol}^\zeta$,  we find $\zeta=0.67\pm0.03$, while \citetalias{macfarlane21} gives $\zeta\sim0.65$. This positive correlation is in line with our current knowledge on galaxy evolution and AGN properties \citep[e.g.][]{jiang07}{}{}, as the gas ensemble within the halo fuels both host galaxy star-forming activity \citep[][]{kennicutt98b} and the accretion activity around black holes. Therefore, higher quasar optical luminosities would indicate a more abundant gas reservoir in the galaxy haloes, which is tied to more intense star formation in host galaxies \citep[e.g.][and references therein]{koss21}{}{}. Meanwhile, we emphasize that despite the increase in the fraction of high jet power sources at higher optical luminosities, quasars spanning the full range of the distribution of jet powers are seen across the whole optical luminosity space — the increase in $\eta$ is the result of scaling the full power-law distribution to higher $\phi$ values in the overall probability distribution. \par

At fixed $M_i$, our modelled host galaxy SFR also shows a strong increase with redshift out to $z=2.5$, which is in line with previous studies on quasar host galaxies \citep[e.g.][]{bonfield11, rosario12, mullaney12, harrison12}. Note that while the host galaxy SFR increases with both quasar bolometric luminosity and redshift, the effect of Malmquist bias is minimal in our fitting process. This is because our fitting approach uses quasar samples within a small range of bolometric luminosity and redshift, therefore our dependencies on luminosity are derived for bins at fixed redshifts and vice versa. While the \citetalias{macfarlane21} correlation showed hints of a turnover beyond $z=2.5$, there were not enough redshift coverage and too few $M_i$ bins to properly characterise it; alternatively, our result shows that despite a flattening in the correlation, the SFR continues to increase with redshift out to $z\sim4$. The reason behind such an increase might be connected to the high prevalence of merger-induced starburst activities at earlier cosmic times, which is often linked to the triggering mechanism of powerful quasars \citep[e.g.][]{sanders88,hopkins06}{}{}. \citet{lamastra13} predicted that the percentage of burst-dominated star forming galaxies increases with redshift, from $\leq 0.5\%$ at $z\sim0.1$ to $\sim20\%$ at $z\sim5$, due to merger-induced starbursts. The SFRs of quasar host galaxies that we derive at higher redshifts lie beyond the predicted values of the star-forming main sequence, hence showing an association with starburst galaxies, and thus an enhancement at higher redshifts relative to the global cosmic SFR density. \citet{duncan19} further posed observational constraints on the merger histories up to $z\sim6$, indicating a higher merger fraction at $z>3$, which is consistent with the cosmic epochs when our SFH of AGN host galaxies differentiates from the cosmic SFR density in star-forming galaxies.

On the other hand, the jet activity level shows little sign of evolution with redshift (see middle panel of Figure~\ref{fig:result_full}). There is a small upturn in the lowest redshift bins in the $\eta$ plots, but this is a result of integrating the power-law jet distribution to a fainter limit at low redshifts due to the lower value of $L_\mu$, rather than a true evolutionary trend: the lack of a trend of either $f$ (Figure~\ref{fig:result_color_mcmc}) or $\gamma$ (Figure~\ref{fig:gamma_fit_slice} and~\ref{fig:gamma_fit_full}) with redshift implies that the distribution of jet powers of the quasars is largely unchanged. While the AGN activity and host galaxy star formation are both tied to the gas reservoir within the galactic halo, the lack of redshift dependency suggests that the former is more likely to be dictated by local activities around the galactic core rather than the time evolution of the gas reservoir. \par

It is not possible to provide direct evidence to the mechanism that powers the AGN radio emission within the scope of this work; however, if we assume this emission arise from radio jets, since the level of jet activity links to the jet powering efficiency of the investigated quasar population, we can still speculate on the physical processes behind AGN jet production upon knowing the evolution pattern of parameter $\eta$. Some studies \citep[e.g.][]{blandford82,woo02} back up the claim that black hole spin is related to the efficiency of jet production \citep{wilson95, velzen13}, with more rapidly-spinning black holes producing more powerful radio jets. In this picture a RL/RQ dichotomy exists, since the black hole spins tend to be either high or low depending on their accretion history. On the other hand, the excellent fits to the data for our model, in which there is no dichotomy between RL and RQ quasars but rather a continuous distribution of jet powers from very high ($\sim10^{30}$W~Hz$^{-1}$) down to sub-dominant compared to the radio emission from SF, is inconsistent with that theory: the wide range of jet power distribution requires a corresponding wide range of black hole spin parameters, which is hard to find in simulations \citep[e.g.][]{volonteri13}. Instead, our result tends to favour the alternate theory presented in \citet{sikora13} that the jet power is controlled by the magnetic flux threading a spinning black hole. Their model is able to create radio jets spanning a wide power range from a variety of magnetic black hole accretion flows, including “magnetically-choked accretion flows” for high-luminosity jets and magnetic field fluctuation in the coronae above thinner accretion disks for low and intermediate power jets. They therefore predict an increase in jet power with increasing black hole accretion rate; both predictions are in line with our results. Note that the discussions above assume a moderate range of black hole masses, while more massive black holes can also produce a stronger magnetic field that leads to increased jet power. We aim to break the degeneracy between black hole accretion rate and black hole mass in our future work. \par

As for the two remaining parameters, we are able to obtain fitting results on $\sigma_\mu$ through the Bayesian scheme and $\gamma$ through examining the radio-loud sources. Our results for $\sigma_\mu$ showed little sign of variation with either $M_i$ or $z$ (see Table~\ref{tab:result}), while a similar situation with $\gamma$ has already been discussed earlier and is shown in Figure~\ref{fig:gamma_fit_slice}. Both results agree with the analysis in \citetalias{macfarlane21}. \par

\section{Link between dust attenuation and radio emission}
\label{sec:colour}

Thanks to the LoTSS DR2 data, we now have additional sample statistics available within the most populated regions of the parameter space. With these extra measurements, we can use the method proposed in this work to study the variance of host galaxy SFR and jet power distribution separately with any measured physical quantity for the quasars, by further splitting the parameter space in our fitting. In particular, in order to definitively identify the origin of the extra radio emission associated with red QSOs discussed in Section~\ref{sec:intro}, we split our samples by optical colour. \par

In this work, instead of defining a subsample of red quasars based on a redshift-dependent percentile cut-off in observed $g-i$ colour \citep[][]{klindt19} or $\Delta E(B-V)$ colour excess \citep[deviation from the average $E(B-V)$ value;][]{glikman22}{}{}, we investigated the colour-related variations across the full range of colour excess; we then discuss our results in the context of previous similar studies on the properties of red/blue quasars. This allows us to speculate into the continuous evolutionary trend of the relevant parameters, rather than only comparing the differences between two populations. As discussed in Section~\ref{sec:data_properties}, we focus on $E(B-V)$ instead of observed $g-i$ colour, as we assume that the redder colour in rQSOs is due to dust extinction only, which has been verified in \citet{fawcett22}. By using $E(B-V$), we can therefore better compare between different bins without trends between different $M_i-z$ bins introducing any biases. \par

\begin{figure}
	\includegraphics[width=\columnwidth]{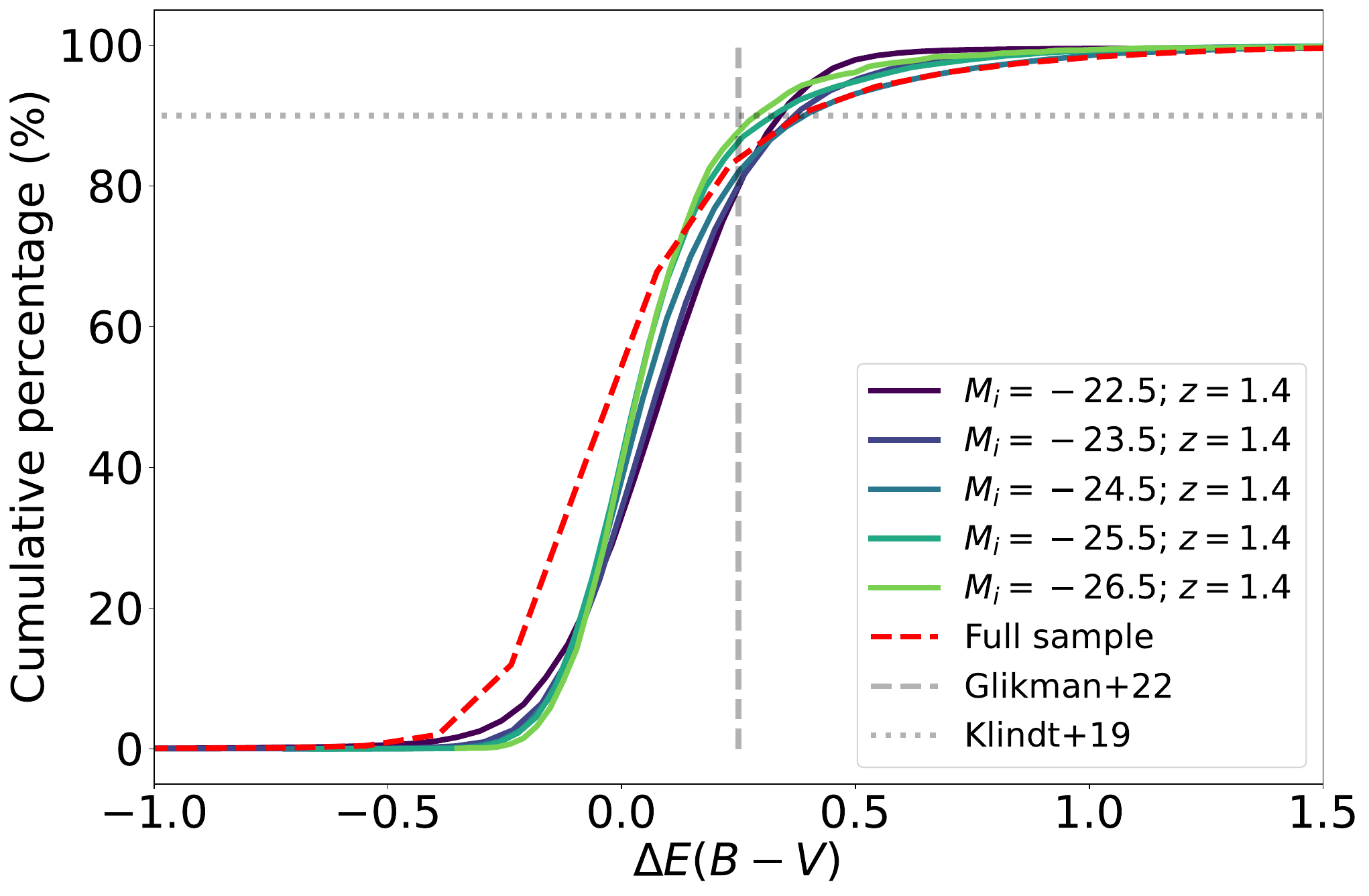}
    \caption{The cumulative distribution of $\Delta E(B-V)$ (deviation in $E(B-V)$ from the average value) in the most populated redshift range ($1.2<z<1.6$). The $M_i$ and $z$ values in the legend are the mid-points of the corresponding $M_i-z$ bins. The observed $g-i$ colour has been corrected to the dust-independent $E(B-V)$ colour excess using the correction method presented in Section~\ref{sec:data_properties}. The grey dotted lines show the definition of red QSOs in \citet{glikman22} ($\Delta E(B-V)>0.25$) and \citet{klindt19} (upper 10\% of the colour distribution). The two criteria intersect close to the cumulative colour distribution curve, suggesting that our sample shows a similar intrinsic colour distribution with both works.}
    \label{fig:color_dist}
\end{figure}

Figure~\ref{fig:color_dist} shows the cumulative distribution of the redshift-corrected $\Delta E(B-V)$ (i.e. the difference of $E(B-V)$ from the average value) in the most densely populated redshift bin ($z=1.4$) for the parameter space explored in this section. The grey dotted line and dashed line compare the selection criteria for rQSOs in \citet[][upper 10\% of colour distribution]{klindt19} to the criteria in \citet[][$\Delta E(B-V)>0.25$]{glikman22} . Both selections broadly agree with each other across the entire luminosity range, suggesting our sample shows a similar intrinsic colour distribution with the previous works.\par

To define a continuous trend of parameter variation, we binned each grid in the $M_i-z$ space into 10 sub-grids based on the colour percentile. Note that we only picked grids with more than 10,000 sources in this process so that each sub-grid contains at least 1,000 sources, which is consistent with the required minimum source number to obtain good fits stated in Appendix~\ref{sec:mocktest}. \par

\begin{figure}
	\includegraphics[width=\columnwidth]{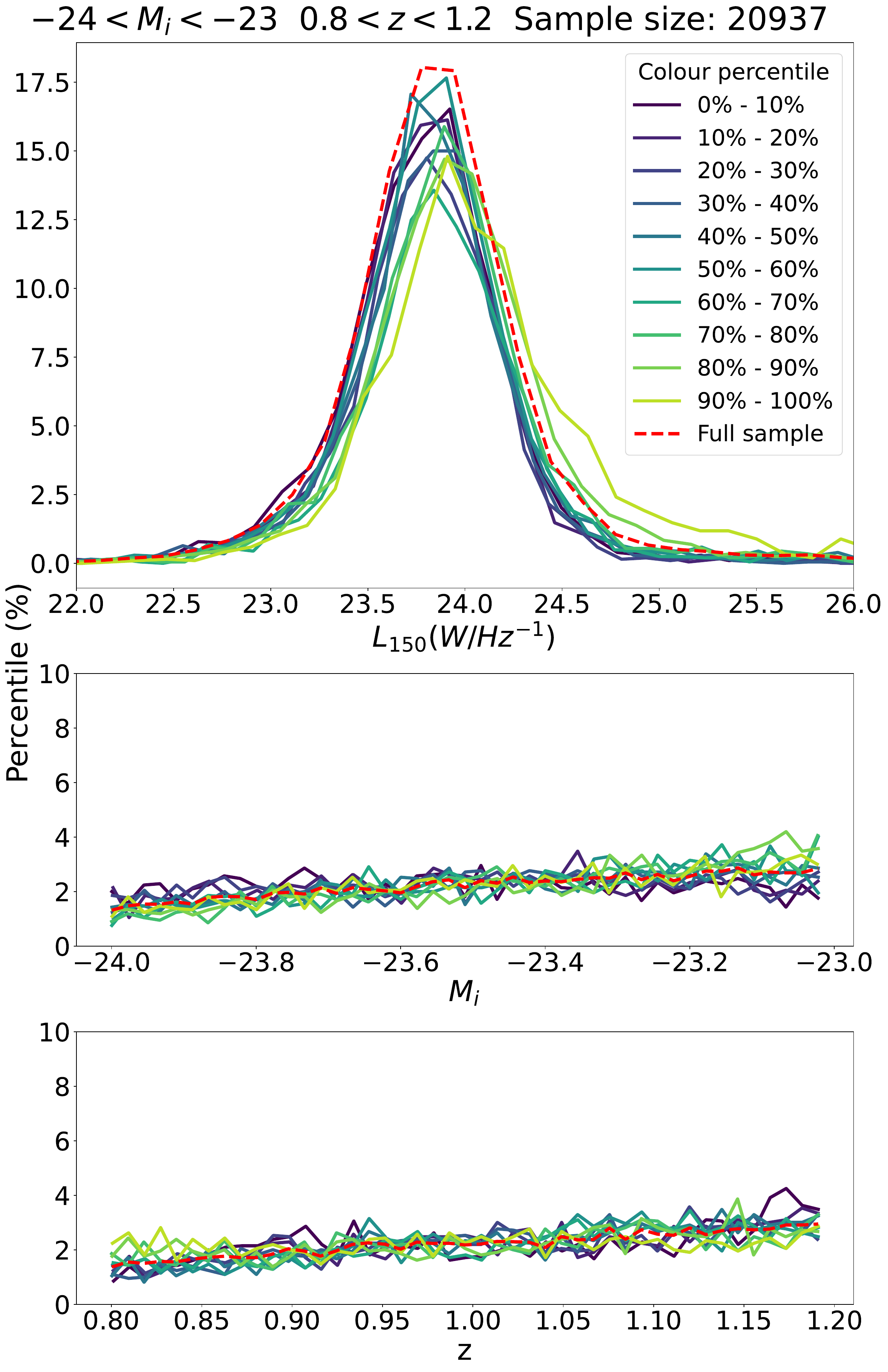}
    \caption{Distributions of sample properties within a representative $M_i$-$z$ grid, characterised by radio luminosity ($L_{150}$), $i$-band magnitude ($M_i$) and redshift ($z$), while separated by 10-percentile colour sub-grids. The distributions of $M_i$ and $z$ show no significant variation across the sub-grids, indicating a uniform sample property among different colour percentiles. Such uniformity enables comparison of fitted values across different colour percentile samples. The distributions of $L_{150}$, on the other hand, show an extended radio-bright wing in the reddest samples, while the locations of the star-forming peak (as defined in Figure~\ref{fig:pdf}) remain unchanged across different colour percentiles. A detailed analysis is presented in Figure~\ref{fig:result_color_mcmc}. Such distributions are seen in all $M_i$-$z$ grids used for quasar colour analysis.}
    \label{fig:sample_dist_color}
\end{figure}

We characterise the samples in each sub-grid by their radio luminosity ($L_{150}$), $i$-band magnitude ($M_i$) and redshift ($z$). The distributions of these physical properties across different colour sub-grids are presented in Figure~\ref{fig:sample_dist_color}, using the quasar samples from the $-24<M_i<-23$, $0.8<z<1.2$ grid. The similarity of both $M_i$ and $z$ distributions across the colour sub-grids indicates there are no systematic biases when studying the evolution of SF and AGN activities with quasar colour percentiles. On the other hand, the distributions of $L_{150}$ differ in the reddest populations (top, i.e. 30\% colour percentile) from the rest, in terms of an extended wing on the radio-bright end. This trend is seen across all grids inspected in our study (see Figure~\ref{fig:sample_dist_color_full}).\par

\begin{figure*}
	\includegraphics[width=\textwidth]{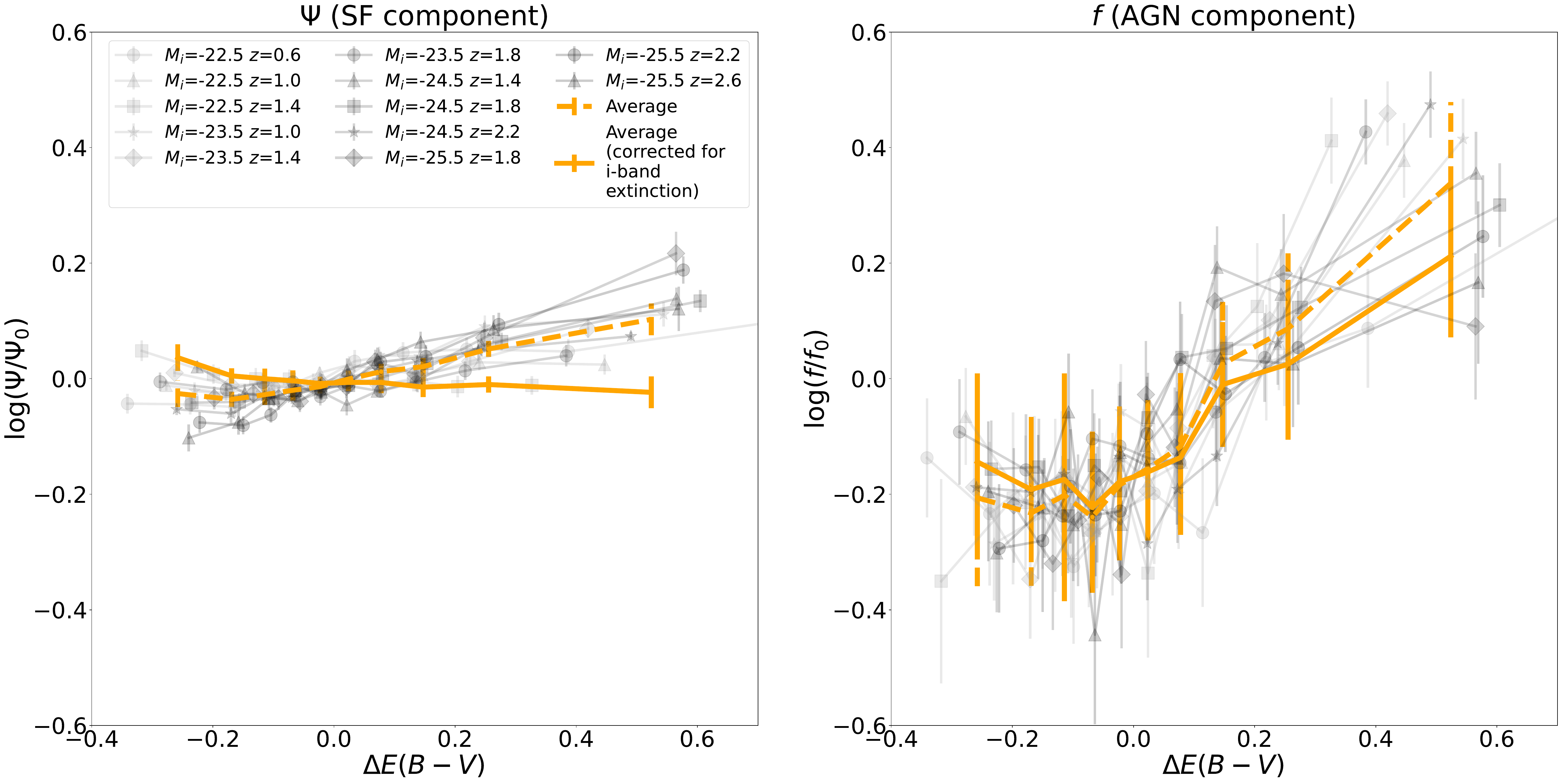}
    \caption{The relative variation of model parameters ($\Psi/\Psi_0$, $f/f_0$) plotted against the deviation in quasar colour excess from the average value ($\Delta E(B-V)$), where $\Psi$, $f$ are the best-fit parameter values within each colour percentile bin, and $\Psi_0$, $f_0$ are the best-fit value using the entire $M_i$-$z$ grid. Within each $M_i$-$z$ grid, the quasars are binned into 10 bins based on the number percentile of $E(B-V)$ colour. Grey lines indicate the trend in each grid cell, whereas the orange dashed line shows the average variance of all investigated grid cells. The left panel shows the variation in the mean host galaxy SFR ($\Psi$), which appears to be a slight increase in $\Psi$ for redder colours (orange dashed line) but shows no variation with $\Delta E(B-V)$ once the effect of dust extinction biasing the $M_i$ values has been removed (orange solid line; see text). The right panel displays the variation in radio jet normalisation ($f$), while the dashed and solid lines represent the variation before and after correcting for dust extinction effect, respectively. As the quasar colour becomes redder, we see a significant increase in the relative jet intensity, showing that the cause of the radio excess in red quasars is indeed due to increased AGN (jet) activity rather than the host galaxy SF.}
    \label{fig:result_color_mcmc}
\end{figure*}

The difference in the distribution of radio powers can thus be accurately modelled by our Bayesian approach. Figure~\ref{fig:result_color_mcmc} shows the relative variation of host galaxy SF activity (characterised by the mean radio SFR $\Psi$) and AGN activity (characterised by the jet normalisation parameter $f$) with the $E(B-V)$ colour percentile. Here, the term 'relative variation' is defined as the ratio between the fitted value within each colour sub-sample ($\Psi$, $f$) and the fitted value using the entire population within the parent grid cell prior to splitting by colour ($\Psi_0$, $f_0$). \par

Since our model has the ability to separate the contribution from SF and AGN in the radio flux density distribution, we can demonstrate that the AGN activity is the main driver behind the reddening of quasars as it strongly correlates with the optical colour (right panel); quasars with the reddest colour also have the highest jet fraction. The host galaxy SF, on the other hand, shows a much weaker correlation with optical colour (left panel). Comparing the correlations across different bins (grey lines), we see little sign of redshift evolution, which suggests that rQSOs are not associated with the star forming activity of their host galaxies, but more likely with the AGN evolutionary stages. The difference in the redness of quasars is therefore related to the dust content surrounding the AGN rather than the dustiness of the host galaxy. Note that our colour-split samples only cover three $M_i-z$ grids beyond the cosmic star formation peak at $z\sim2$; therefore deeper radio observations are needed to draw a complete census across cosmic times. Using additional multiwavelength data in the smaller LoTSS Deep Field DR1 \citep[][]{lotssdeepdr1, lotssdeepen1}  coverage instead, \citet{calistrorivera23} developed an independent method to separate AGN and SF components in quasar SEDs and came to a similar conclusion that radio emission in rQSOs comes almost exclusively from AGN. Furthermore, as the host galaxy SF also correlates with the absolute i-band magnitude (Figure~\ref{fig:result_full}), this weak trend between SF and optical colour may as well be the side effect of extinction. \par

Extinction will result in the measured $i$-band magnitude being fainter than it should be due to the dust, especially for quasars in the redder bins. Since correction based on the extinction law will again increase the measurement of $M_i$, the corrected values for $L_\mathrm{bol}$ will increase in higher bins; as a result, the expected SFR will also be higher due to the correlation between $\Psi$ and $M_i$ in Figure~\ref{fig:result_full}, which may produce an artificial trend shown by the orange dashed line in Figure~\ref{fig:result_color_mcmc}. To test this, we deduced $\Delta \log\Psi=a\Delta  M_i$ from the fitted value, where $a$ is the slope inferred from the $\textrm{SFR}- L_{bol}$ correlation in Section~\ref{sec:result_evo}, and $\Delta M_i$ is the correction in the measured $i$-band magnitude based on the extinction law and the given $E(B-V)$. The weak correlation observed between fitted $\Psi$ and $\Delta E(B-V)$ is removed after the correction (solid line), confirming that this correlation is merely an artefact of the dust reddening in the photometric bands. This result shows that the SF contribution to the radio emission remains unchanged between red and blue QSOs of the same bolometric luminosity. We have taken a similar approach to the $\log f-\Delta E(B-V)$ relation using a slope inferred from the $f-L_{bol}$ correlation in Section~\ref{sec:result_evo}, and the corrected relation is shown as an orange solid line in the right panel. The corrected result still shows a strong positive correlation between AGN activity and optical reddening, and we therefore rule out the possibility that this correlation arises from the side effect of dust extinction. \par

\begin{figure}
	\includegraphics[width=\columnwidth]{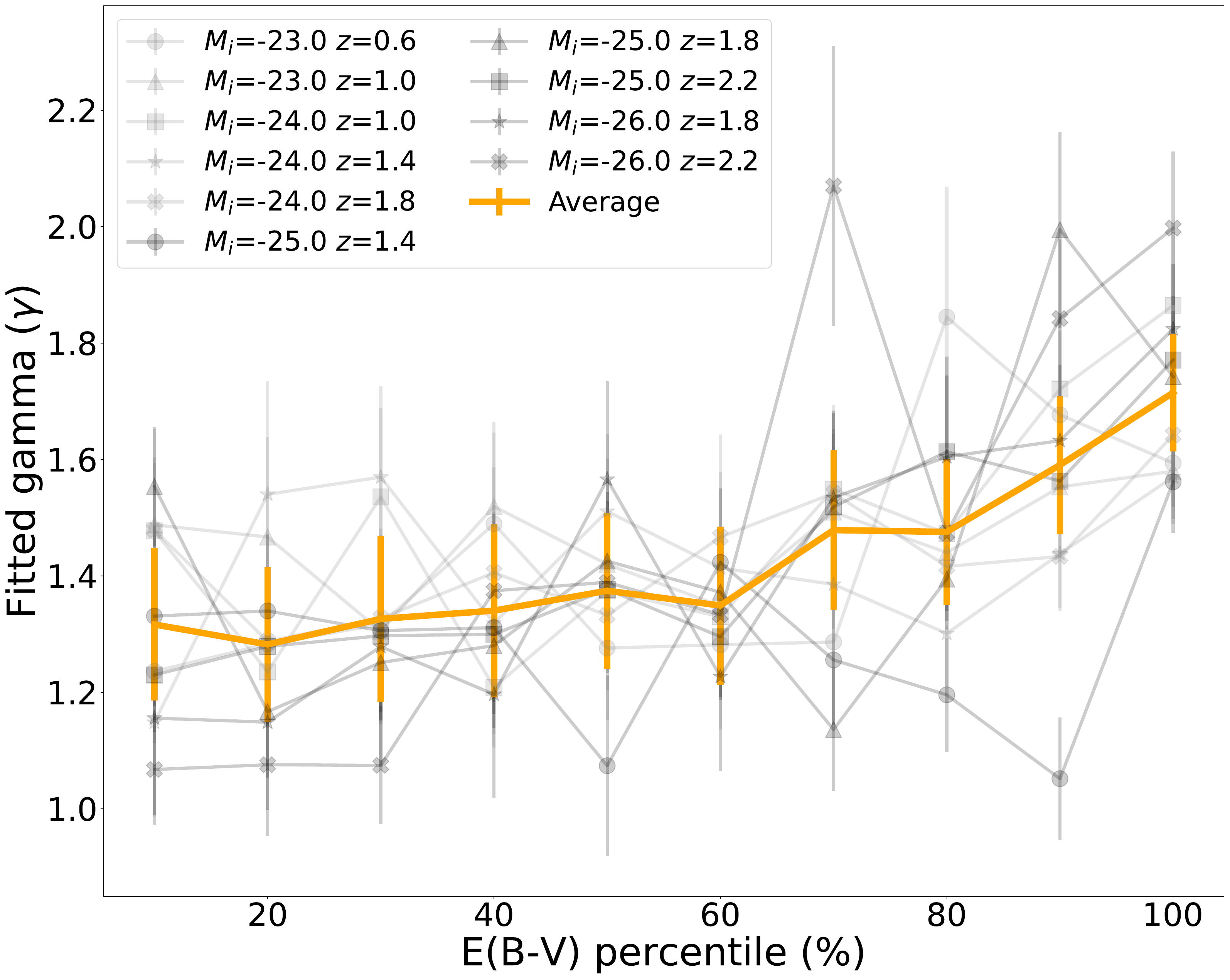}
    \caption{The variation of $\gamma$ with the $E(B-V)$ percentile, where the grey lines show the fitted value within each grid bin, and the orange line shows the averaged value across the entire parameter space. The best-fit values of $\gamma$ show signs of increase at redder colours, especially within the top 10\% of $E(B-V)$ distribution. This indicates more radio-faint rQSOs have been enhanced than radio-bright rQSOs, therefore causing the rise in the power-law slope at the radio-bright end of the radio flux density distribution.}
    \label{fig:result_color_gamma}
\end{figure}

Recently, \citet{fawcett22} argued that radio excess in rQSOs occurs primarily in quasars with lower radio powers, likely due to the interaction between weak/intermediate jets and the opaque interstellar medium/circumnuclear environment in rQSOs. Our model allows us to speculate about the details of this process by exploring whether $\gamma$ is changing. Figure~\ref{fig:result_color_gamma} shows the variation in fitted $\gamma$ values across different colour percentiles, using the fitting method described in Section~\ref{sec:priortest}. Our results reveal a universal increase in fitted $\gamma$ values for rQSOs, which indicates that more radio-faint quasars have been enhanced than radio-bright quasars in the reddest bins, thus causing the rise in the slope of flux density distribution in the radio-bright end. We therefore confirm previous literature results \citep[][]{klindt19,rosario20,fawcett20,fawcett22,calistrorivera23}, and conclude that the change in evolutionary phase that caused the reddening of rQSOs takes place mostly in radio-quiet and radio-intermediate quasars. \par

It is also worth noting that our results do not rule out the contribution of wind shocks in the radio enhancement of red quasars, as investigated in Petley et al. (\textit{in prep}) based on a sample of broad absorption line quasars (BALQSOs). While our model can effectively rule out the host galaxy contribution in the radio enhancement, the AGN component in our model consists of any radio emission generated in BH activity that does not follow a Gaussian distribution centred at the average radio SFR, including jet and wind activities. The steepening of $\gamma$ could also be indicative of faint AGN radio emission other than weak jet (e.g. wind shock) being enhanced in the red quasars. If the radio luminosity function of that other component becomes reasonably well understood in the future, the methodology developed in this work can easily be expanded to explicitly include a third component (e.g. merger history, wind shocks, etc.).

\section{Conclusions}
\label{sec:conclusion}

In this work, we adopted the assumption made in \citetalias{macfarlane21} and developed a fully Bayesian two-component model (see Figure~\ref{fig:pdf}) that statistically disentangles the host galaxy SF and AGN jet contributions to the quasar radio emission, using the quasar radio flux density distribution as input. We assumed a log-Gaussian distribution for the SF flux densities with two parameters $L_\mu$ (mean radio SF luminosity) and $\sigma_\mu$ (Gaussian scatter of radio SF luminosity), and a single power-law distribution for the jet flux densities with an additional two parameters $f$ (or $\eta$; the jet power normalisation factor) and $\gamma$ (power-law slope). To investigate the variation of best-fit values of our model parameters with a number of factors including bolometric luminosity ($L_\mathrm{bol}$), redshift ($z$) and optical reddening (characterised by the colour excess $E(B-V)$, we binned our samples into grid bins based on their location in the parameter space of $M_i$, $z$ and colour, before fitting the quasar radio flux density distributions within each grid bin against our model. \par

After various analyses, we have reached the following conclusions:

\begin{enumerate}
    \item Our model confirms the main results of \citetalias{macfarlane21} but with a higher accuracy and a wider coverage of the parameter space, probing sources at a magnitude deeper and with a redshift out to $z\sim4$. We obtained good fits using our two-component model across the entire parameter space, which argues against an RL/RQ dichotomy. \par

    \item The host galaxy SFR ($\Psi$) and jet activity ($\eta$) share a positive correlation with $M_i$, which can be associated with the black hole accretion rate. While the host galaxy SFR also positively correlates with redshift, the AGN jet activity experiences little evolution across cosmic time (Figure~\ref{fig:result_full}). Our result supports the speculation in \citetalias{macfarlane21} that the overall gas fraction and dynamical time within the halo will affect both star formation and black hole accretion; the fact the host galaxy SFR continues to grow when $z>2.5$ goes beyond the \citetalias{macfarlane21} result, as they lacked sufficient source to probe beyond the redshift limit, and is different from the cosmic star formation rate density measured in star-forming galaxies \citep[e.g.][]{madau14}{}{}. This trend can be explained by the high prevalence of galaxy mergers in earlier cosmic times, which are likely to be responsible for the triggering mechanism of both powerful quasars and starburst activities.
    %Further study is required to investigate the possible influence of enhanced starburst activities within these AGN host galaxies. \par

    \item  The host galaxy SFR is found to have little impact on the excessive radio emission in red quasars (rQSOs), while the jet activity plays a significant role as it is found to be increasing at all colours redder than the average quasar colour (Figure~\ref{fig:result_color_mcmc}). This provides direct evidence to the evolutionary model that explains the radio enhancement in rQSOs, and rules out the possibilities of SF influence. The slope of the radio-bright end of the flux density distribution shows signs of positive correlation with the quasar colour (Figure~\ref{fig:result_color_gamma}), which indicates rQSOs with weak or intermediate jet activities are more likely to experience a radio enhancement.
\end{enumerate}

This work shows the flexibility of our model that will enable us to investigate the role of host galaxy and AGN separately in terms of the correlation between quasar radio emission and various other physical parameters, including the black hole mass, clustering environment, AGN winds and outflows, etc. We plan to extend our method to examine the correlations stated above using the newest LoTSS and SDSS survey data. Future spectroscopic surveys including WEAVE-LOFAR \citep{weave} will provide more robust measurements to the continuum and emission line properties of LoTSS-detected quasars, thus giving a deeper insight into the details of the accretion processes, outflow, circumnuclear environments and dust composition of quasars. Other surveys such as DESI \citep{desi} and WEAVE-QSO \citep{weaveqso} will also further expand the samples of optically selected quasars. Combining the techniques and new observational data, we are hopeful of understanding the sources of radio emission within quasars and their relative contributions.

\section*{Acknowledgements}

We thank the anonymous referee for the useful comments that helped shape the final version of this paper. BY would like to thank for the support from the University of Edinburgh and Leiden Observatory through the Edinburgh-Leiden joint studentship. PNB is grateful for support from the UK STFC via grant ST/V000594/1. KJD acknowledges funding from the European Union's Horizon 2020 research and innovation programme under the Marie Sk\l{}odowska-Curie grant agreement No. 892117 (HIZRAD) and support from the STFC through an Ernest Rutherford Fellowship (grant number ST/W003120/1). This work was supported by the Medical Research Council [MR/T042842/1]. JP acknowledges support for their PhD studentship from grants ST/T506047/1 and ST/V506643/1. IP acknowledges support from INAF under the Large Grant 2022 funding scheme (project “MeerKAT and LOFAR Team up: a Unique Radio Window on Galaxy/AGN co-Evolution”. DJBS acknowledges support from the UK STFC [ST/V000624/1].\par

For the purpose of open access, the author has applied a Creative Commons Attribution (CC BY) licence to any Author Accepted Manuscript version arising from this submission. \par

LOFAR data products were provided by the LOFAR Surveys Key Science project (LSKSP; \url{https://lofar-surveys.org/}) and were derived from observations with the International LOFAR Telescope (ILT). LOFAR \citep[][]{lofar}{}{} is the Low Frequency Array designed and constructed by ASTRON. It has observing, data processing, and data storage facilities in several countries, which are owned by various parties (each with their own funding sources), and which are collectively operated by the ILT foundation under a joint scientific policy. The efforts of the LSKSP have benefited from funding from the European Research Council, NOVA, NWO, CNRS-INSU, the SURF Co-operative, the UK Science and Technology Funding Council and the J\"ulich Supercomputing Centre. This research made use of the University of Hertfordshire high-performance computing facility and the LOFAR-UK computing facility located at the University of Hertfordshire and supported by STFC [ST/P000096/1].\par

Funding for the Sloan Digital Sky Survey IV has been provided by the Alfred P. Sloan Foundation, the U.S. Department of Energy Office of Science, and the Participating Institutions. \par

SDSS-IV acknowledges support and resources from the Center for High Performance Computing at the University of Utah. The SDSS website is \url{www.sdss4.org}. \par

SDSS-IV is managed by the Astrophysical Research Consortium for the Participating Institutions of the SDSS Collaboration including the Brazilian Participation Group, the Carnegie Institution for Science, Carnegie Mellon University, Center for Astrophysics | Harvard \& Smithsonian, the Chilean Participation Group, the French Participation Group, Instituto de Astrof\'isica de Canarias, The Johns Hopkins University, Kavli Institute for the Physics and Mathematics of the Universe (IPMU) / University of Tokyo, the Korean Participation Group, Lawrence Berkeley National Laboratory, Leibniz Institut f\"ur Astrophysik Potsdam (AIP),  Max-Planck-Institut f\"ur Astronomie (MPIA Heidelberg), Max-Planck-Institut f\"ur Astrophysik (MPA Garching), Max-Planck-Institut f\"ur Extraterrestrische Physik (MPE), National Astronomical Observatories of China, New Mexico State University, New York University, University of Notre Dame, Observat\'ario Nacional / MCTI, The Ohio State University, Pennsylvania State University, Shanghai Astronomical Observatory, United Kingdom Participation Group, Universidad Nacional Aut\'onoma de M\'exico, University of Arizona, University of Colorado Boulder, University of Oxford, University of Portsmouth, University of Utah, University of Virginia, University of Washington, University of Wisconsin, Vanderbilt University, and Yale University. \par

\section*{Data Availability}

The datasets used in this paper were derived from sources in the public domain: the LOFAR Two-Metre Sky Surveys (\url{www.lofar-surveys.org}) and the Sloan Digital Sky Survey (\url{www.sdss.org}).



\bibliographystyle{mnras}
\bibliography{refjournal} % if your bibtex file is called example.bib

%\begin{thebibliography}{99}
%\end{thebibliography}

%%%%%%%%%%%%%%%%%%%%%%%%%%%%%%%%%%%%%%%%%%%%%%%%%%

%%%%%%%%%%%%%%%%% APPENDICES %%%%%%%%%%%%%%%%%%%%%

\appendix

\section{Testing the parametric model using mock data input}

\subsection{Binning}
\label{sec:mocktest}
Since the physical properties of quasars span over a large range of parameter space, we need to bin our sample quasars into small areas within the parameter space. This results in different numbers of samples being put into the fitting process, and therefore impact the quality of our fit. Having too few sources within each bin could result in small number statistics and thus affect the accuracy of our model; on the other hand, having too many sources within each bin could affect the resolution of our parameter space. Different values of parameters also lead to different levels of model degeneracy, thus compromising the fitting quality.\par

We used mock quasar radio flux density distributions generated from the probability density function described in \ref{sec:model_twocomp}, with a wide range of input parameter values and a variety of sample sizes, to understand how the number of input sources affects the fitting quality and in which parts of parameter space can we reliably measure the values. We assumed a uniform redshift within the bin based on the redshift distribution pattern shown in Figure~\ref{fig:sample_dist_color}. The fitting quality is characterised by the relative difference between the input parameter used to generate the mock samples and the best-fit result from our algorithm: 

\begin{equation}
    \delta=\frac{p_\textrm{best-fit}-p_\textrm{input}}{p_\textrm{input}}.
    \label{eq:A1.1}
\end{equation}

\begin{figure}
	\includegraphics[width=\columnwidth]{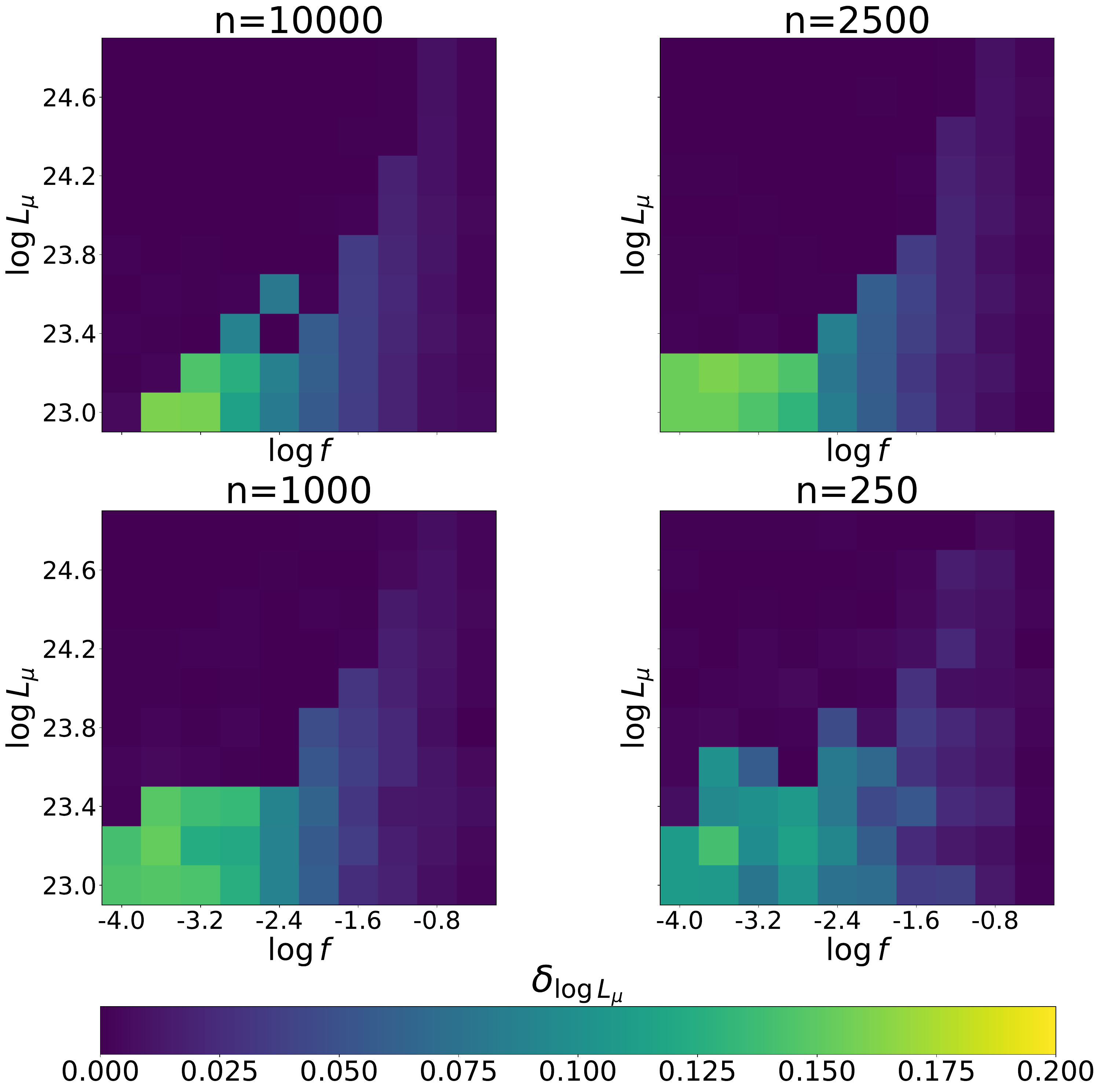}
    \caption{Relative difference between input and best-fit values of $L_\mu$ (mean host galaxy radio SF luminosity). Note that here the relative difference is defined by Equation~\ref{eq:A1.1}. We adopted a simple box prior to $\sigma_\mu$ and a sharp Gaussian prior to $\gamma$ centred on $\gamma=1.5$, as described in Section~\ref{sec:model_fitting}. We thus investigate the dependency on the input values of the remaining two parameters: $L_\mu$ and $f$ (fraction of jet-dominated quasars). For $L_\mu$, the best-fit result reaches maximum difference in the bottom left corner where both SF and jet components are weak, therefore making it hard to separate the two components. As the jet component grows stronger (or the SF luminosity increases), the SF contribution becomes easier to separate from the entire distribution. The pattern remains valid for all sample numbers, while the lower limit of the parameter space with good fits corresponds to 5 times the stacked noise level. Lower source counts lead to higher stacked noise levels, thus limiting the suitable parameter space.}
    \label{fig:mock_logl}
\end{figure}

\begin{figure}
	\includegraphics[width=\columnwidth]{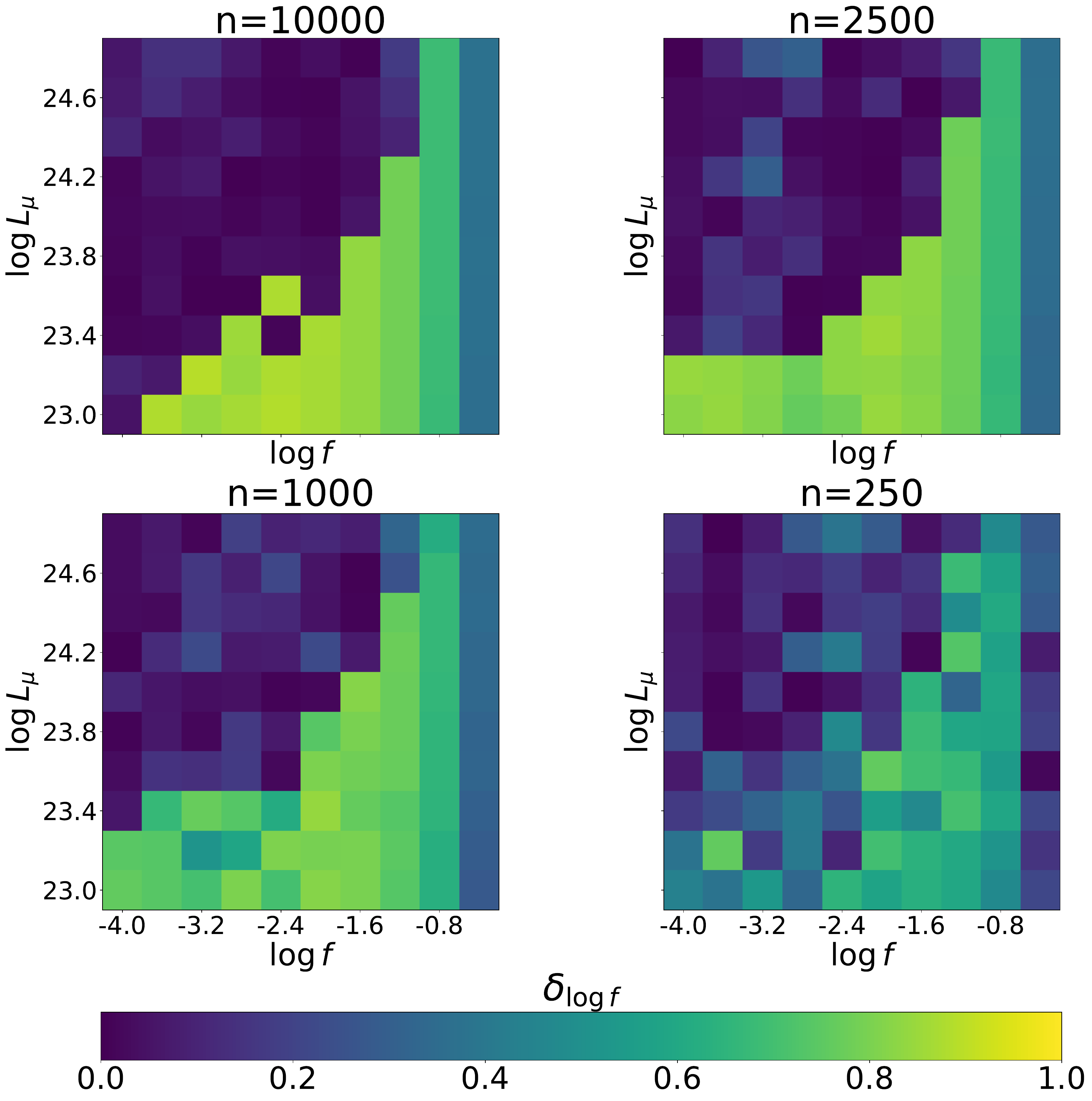}
    \caption{Relative difference between input and best-fit values of $f$ (fraction of jet-dominated quasars). While the pattern and the limit for suitable parameter space remain similar to Figure~\ref{fig:mock_logl}, the best-fit values of $f$ is more sensitive to input, and the uncertainties on the derived values for $f$ increase in the regions of parameter space where the input jet-dominated quasar fraction is high or the SFR is low.}
    \label{fig:mock_logf}
\end{figure}

Figure \ref{fig:mock_logl} and \ref{fig:mock_logf} shows the fitting quality (characterised by $\delta_{\log L_\mu}$ and $\delta_{\log f}$) of $L_\mu$ and $f$ under different model parameters (characterised by the $\log L_\mu$ - $\log f$ parameter space) and different number of fitted sources. We targeted these two parameters for investigation because they are pivotal to the model construction (see Section \ref{sec:model_twocomp}), and have important roles in explaining the physical process of quasars ($L_\mu$ translates to the mean luminosity of host galaxy star formation and $f$ to the fraction of jet-dominated quasars). For the rest of the parameters, we adopted the priors described in Section~\ref{sec:model_fitting} for consistency.

Both figures show a minor trend between the fitting quality and the number of sources included. As the number of sources decreased from 10,000 to 1000, the lower limit of intrinsic $L_\mu$ required for a good fit increased, in a trend corresponding to the increase of stacked noise level of simulated sources. The poorest fitting quality of $L_\mu$ occurs at the faint end of both SF and jet emission, and gradually improves with increasing intrinsic SF and jet luminosity - note that there is no significant difference in the fitting quality once the SF luminosity reaches 10 times the stacked noise level, although higher jet fraction leads to slightly worse results in $L_\mu$ under the same intrinsic SF luminosity. The fitting quality of $f$, on the other hand, correlates more with the intrinsic jet fraction compared to the situation in $L_\mu$. While the limit of intrinsic SF is still governed by the stacked noise level, an additional limit to intrinsic jet power is found in the parameter space, as the fitting quality for $f$ worsened in the high-SF, high-jet power regime. This is mostly due to the confusion between the bright end of the SF distribution and the faint end of the jet distribution, when the increased mean SF luminosity and jet power normalisation moved these two populations into the same space in the radio flux density distribution. The fitting quality within the parameter space for good fits also slightly deteriorated for $f$, when compared to $L_\mu$. \par

When the source count falls down to 250, the fitting quality continues to deteriorate, so that we cannot determine a clear boundary within our parameter space for guaranteed good fits. Therefore, we chose 1,000 sources as the minimum number of quasars required for a good fit. We selected the target grid bins for analysis in the $M_i-z$ plane based on this criterion. For colour-dependent studies, the minimum number is set to 10,000 sources since we are binning each grid bin into 10 sub-grids based on the colour percentile, so that each sub-grid contains at least 1,000 sources.

\subsection{Priors}
\label{sec:priortest}

\begin{landscape}
    \begin{figure}
        \includegraphics[width=\columnwidth]{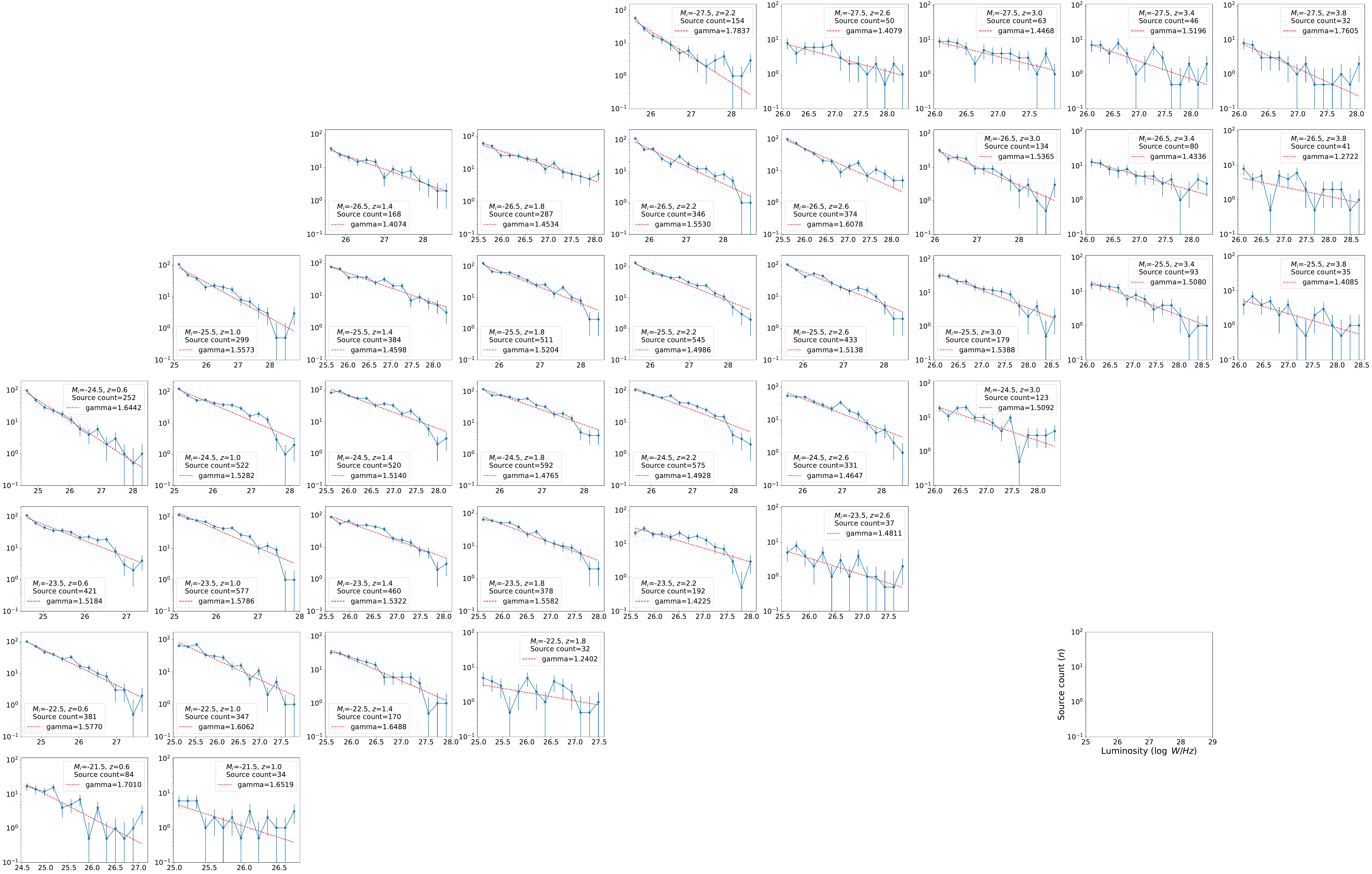}
        \caption{The bright end of the quasar radio luminosity function in each grid bin, used for determining the prior of $\gamma$ in the fitting. We picked the faint cut-off for the radio-bright luminosity function by visual inspection so that the cut-off values lie above the $10\sigma$ flux density limits within each bin and are at least 1 dex above the estimated $L_\mu$, therefore the selected distribution only shows the single power-law feature in the $\log n-\log L$ plane and the included sources can be treated as jet-dominated. The number count within each luminosity bin is shown as blue dots, while the uncertainty is determined by the Poisson error within each bin. We fit a single power-law model ($n(L)\propto L^{-\gamma}$) to the radio-bright luminosity functions in all grids (red dotted line), and the fitted values for the power-law slope $\gamma$ are highlighted in the legend. Our single power-law model gives a good fit across all grid bins within the parameter space and further proves the validity of our model regarding the jet components. The fitted values of $\gamma$ lie around $\gamma\approx1.5$ and show little dependence on redshift or optical magnitude (see Figure~\ref{fig:gamma_fit_slice}). The informed prior of $\gamma$ is therefore defined as a Gaussian distribution centered at $\gamma_0=1.5$ with a narrow scatter of $\sigma_\gamma=0.05$ (see Equation~\ref{eq:3.2.14}).}
        \label{fig:gamma_fit_full}
    \end{figure}
\end{landscape}

Following the discussion in Section~\ref{sec:model_fitting}, we present some detailed results used to determine the prior on $\gamma$ in order to eliminate the degeneracy between $\gamma$ and $\sigma_\mu/\sigma_\Psi$ in our fitting. Figure~\ref{fig:gamma_fit_full} shows the selection of the bright-end quasar radio luminosity function fits used to determine the $\gamma$ prior. The determined values of $\gamma$, together with their uncertainties, are given in the final two columns of Table~\ref{tab:result}.\par

\bsp	% typesetting comment

\begin{landscape}
    \begin{table}
        \caption{The variation of best-fit parameter values (for $\gamma$ we show the best-fit values obtained in Section~\ref{sec:priortest}) and uncertainties within the $M_i$-$z$ plane for the full sample, together with the lower luminosity cut-off for determining the radio-loud luminosity function defined in Appendix~\ref{sec:priortest} ($L_{\mathrm{min}}$). We also included the black hole accretion rate ($\log \Dot{M}_{\mathrm{BH}}$) for reference, which is calculated from Equation 2 in \citetalias{macfarlane21}.}
        \label{tab:result}
        \begin{tabular}{cccccccccccccccc}
            \hline
            $M_i$ & $z$ & $\log \Dot{M}_{\mathrm{BH}}$ & $\log L_{\mu}$ & $\Delta\log L_{\mu}$ & $\log\Psi$ & $\Delta\log\Psi$ & $\sigma_\Psi$ & $\Delta\sigma_\Psi$ & $\log f$ & $\Delta\log f$ & $\log\eta$ & $\Delta\log\eta$ & $L_\mathrm{min}$ & $\gamma$ & $\Delta\gamma$\\
            $(z=0)$ & & $[M_{\sun}~\mathrm{yr}^{-1}]$ & $[\mathrm{W~Hz}^{-1}]$ & $[\mathrm{W~Hz}^{-1}]$ & $[M_{\sun}~\mathrm{yr}^{-1}]$ & $[M_{\sun}~\mathrm{yr}^{-1}]$ & $[\mathrm{dex}]$ & $[\mathrm{dex}]$ & & & & & $[\mathrm{W~Hz}^{-1}]$ & & \\
            \hline
            -21.5 & 0.6 & -0.91 & 23.1347 & 0.0025 & 0.9161 & 0.0024 & 0.21 & 0.01 & -2.32 & 0.05 & -0.88 & 0.05 & 24.5 & 1.70 & 0.10\\
            -22.5 & 0.6 & -0.53 & 23.3554 & 0.0012 & 1.1282 & 0.0012 & 0.27 & 0.01 & -1.97 & 0.03 & -0.65 & 0.03 & 24.5 & 1.58 & 0.03\\
            -23.5 & 0.6 & -0.15 & 23.5315 & 0.0217 & 1.2973 & 0.0208 & 0.32 & 0.01 & -1.60 & 0.05 & -0.36 & 0.04 & 24.5 & 1.52 & 0.03\\
            -24.5 & 0.6 & 0.22 & 23.6416 & 0.1301 & 1.4031 & 0.1249 & 0.37 & 0.04 & -1.33 & 0.12 & -0.15 & 0.05 & 24.5 & 1.64 & 0.05\\
            -21.5 & 1.0 & -0.91 & 23.5244 & 0.0169 & 1.2905 & 0.0162 & 0.17 & 0.02 & -2.28 & 0.07 & -1.04 & 0.07 & 25.0 & 1.65 & 0.27\\
            -22.5 & 1.0 & -0.53 & 23.6858 & 0.0010 & 1.4455 & 0.0010 & 0.22 & 0.01 & -2.06 & 0.04 & -0.90 & 0.04 & 25.0 & 1.61 & 0.04\\
            -23.5 & 1.0 & -0.15 & 23.8607 & 0.0210 & 1.6135 & 0.0202 & 0.27 & 0.01 & -1.80 & 0.03 & -0.73 & 0.02 & 25.0 & 1.58 & 0.03\\
            -24.5 & 1.0 & 0.22 & 23.9494 & 0.0216 & 1.6987 & 0.0207 & 0.28 & 0.01 & -1.38 & 0.04 & -0.36 & 0.02 & 25.0 & 1.53 & 0.03\\
            -25.5 & 1.0 & 0.60 & 24.0652 & 0.0908 & 1.8100 & 0.0872 & 0.31 & 0.03 & -1.02 & 0.12 & -0.06 & 0.07 & 25.0 & 1.56 & 0.04\\
            -22.5 & 1.4 & -0.53 & 23.9691 & 0.0196 & 1.7176 & 0.0189 & 0.19 & 0.01 & -2.01 & 0.04 & -0.99 & 0.03 & 25.5 & 1.65 & 0.06\\
            -23.5 & 1.4 & -0.15 & 24.0800 & 0.0203 & 1.8242 & 0.0195 & 0.24 & 0.01 & -1.88 & 0.03 & -0.92 & 0.02 & 25.5 & 1.53 & 0.03\\
            -24.5 & 1.4 & 0.22 & 24.1909 & 0.0210 & 1.9308 & 0.0202 & 0.27 & 0.01 & -1.61 & 0.03 & -0.70 & 0.02 & 25.5 & 1.51 & 0.03\\
            -25.5 & 1.4 & 0.60 & 24.2793 & 0.0215 & 2.0157 & 0.0207 & 0.29 & 0.01 & -1.20 & 0.04 & -0.34 & 0.03 & 25.5 & 1.46 & 0.03\\
            -26.5 & 1.4 & 0.97 & 24.3124 & 0.0427 & 2.0475 & 0.0410 & 0.39 & 0.06 & -0.87 & 0.16 & -0.03 & 0.13 & 25.5 & 1.41 & 0.05\\
            -22.5 & 1.8 & -0.53 & 24.1472 & 0.0153 & 1.8888 & 0.0147 & 0.20 & 0.02 & -2.01 & 0.11 & -1.08 & 0.11 & 25.0 & 1.24 & 0.14\\
            -23.5 & 1.8 & -0.15 & 24.2529 & 0.0179 & 1.9903 & 0.0172 & 0.21 & 0.01 & -1.85 & 0.04 & -0.97 & 0.03 & 25.5 & 1.56 & 0.04\\
            -24.5 & 1.8 & 0.22 & 24.3655 & 0.0201 & 2.0985 & 0.0193 & 0.25 & 0.01 & -1.66 & 0.03 & -0.85 & 0.02 & 25.5 & 1.48 & 0.02\\
            -25.5 & 1.8 & 0.60 & 24.4548 & 0.0212 & 2.1842 & 0.0204 & 0.28 & 0.01 & -1.30 & 0.03 & -0.53 & 0.02 & 25.5 & 1.52 & 0.03\\
            -26.5 & 1.8 & 0.97 & 24.5205 & 0.0224 & 2.2473 & 0.0215 & 0.29 & 0.02 & -0.89 & 0.07 & -0.15 & 0.06 & 25.5 & 1.45 & 0.04\\
            -23.5 & 2.2 & -0.15 & 24.4110 & 0.0017 & 2.1421 & 0.0016 & 0.17 & 0.01 & -1.74 & 0.04 & -0.95 & 0.04 & 25.5 & 1.42 & 0.05\\
            -24.5 & 2.2 & 0.22 & 24.4773 & 0.0012 & 2.2059 & 0.0011 & 0.22 & 0.01 & -1.57 & 0.03 & -0.81 & 0.03 & 25.5 & 1.49 & 0.02\\
            -25.5 & 2.2 & 0.60 & 24.5879 & 0.0008 & 2.3121 & 0.0008 & 0.27 & 0.01 & -1.38 & 0.04 & -0.67 & 0.04 & 25.5 & 1.50 & 0.03\\
            -26.5 & 2.2 & 0.97 & 24.6976 & 0.0031 & 2.4174 & 0.0030 & 0.31 & 0.01 & -1.08 & 0.06 & -0.43 & 0.05 & 25.5 & 1.55 & 0.04\\
            -27.5 & 2.2 & 1.35 & 24.8607 & 0.0648 & 2.5741 & 0.0622 & 0.40 & 0.05 & -0.91 & 0.18 & -0.34 & 0.15 & 25.5 & 1.78 & 0.08\\
            -23.5 & 2.6 & -0.15 & 24.5386 & 0.0178 & 2.2648 & 0.0171 & 0.07 & 0.01 & -1.65 & 0.06 & -0.92 & 0.06 & 25.5 & 1.48 & 0.16\\
            -24.5 & 2.6 & 0.22 & 24.5845 & 0.0193 & 2.3088 & 0.0186 & 0.20 & 0.01 & -1.54 & 0.03 & -0.83 & 0.02 & 25.5 & 1.46 & 0.03\\
            -25.5 & 2.6 & 0.60 & 24.6952 & 0.0200 & 2.4152 & 0.0192 & 0.25 & 0.01 & -1.33 & 0.04 & -0.68 & 0.03 & 25.5 & 1.51 & 0.03\\
            -26.5 & 2.6 & 0.97 & 24.8077 & 0.0023 & 2.5232 & 0.0022 & 0.29 & 0.01 & -1.01 & 0.05 & -0.41 & 0.05 & 25.5 & 1.61 & 0.04\\
            -27.5 & 2.6 & 1.35 & 24.9398 & 0.0434 & 2.6501 & 0.0417 & 0.32 & 0.04 & -0.78 & 0.13 & -0.25 & 0.11 & 26.0 & 1.41 & 0.10\\
            -24.5 & 3.0 & 0.22 & 24.6739 & 0.0200 & 2.3947 & 0.0192 & 0.11 & 0.03 & -1.34 & 0.05 & -0.68 & 0.04 & 26.0 & 1.51 & 0.07\\
            -25.5 & 3.0 & 0.60 & 24.7851 & 0.0020 & 2.5016 & 0.0019 & 0.23 & 0.01 & -1.26 & 0.04 & -0.65 & 0.04 & 26.0 & 1.54 & 0.05\\
            -26.5 & 3.0 & 0.97 & 24.8939 & 0.0212 & 2.6061 & 0.0203 & 0.28 & 0.02 & -1.01 & 0.06 & -0.46 & 0.05 & 26.0 & 1.54 & 0.06\\
            -27.5 & 3.0 & 1.35 & 25.0173 & 0.0609 & 2.7246 & 0.0585 & 0.33 & 0.04 & -0.74 & 0.16 & -0.25 & 0.13 & 26.0 & 1.45 & 0.12\\
            -25.5 & 3.4 & 0.60 & 24.8456 & 0.0199 & 2.5597 & 0.0191 & 0.23 & 0.02 & -1.14 & 0.07 & -0.56 & 0.06 & 26.0 & 1.51 & 0.07\\
            -26.5 & 3.4 & 0.97 & 24.9681 & 0.0175 & 2.6773 & 0.0168 & 0.28 & 0.03 & -1.01 & 0.09 & -0.50 & 0.08 & 26.0 & 1.43 & 0.09\\
            -27.5 & 3.4 & 1.35 & 25.0908 & 0.0480 & 2.7952 & 0.0461 & 0.28 & 0.04 & -0.69 & 0.14 & -0.24 & 0.12 & 26.0 & 1.52 & 0.14\\
            -25.5 & 3.8 & 0.60 & 24.9122 & 0.0384 & 2.6236 & 0.0369 & 0.13 & 0.05 & -0.97 & 0.10 & -0.43 & 0.08 & 26.0 & 1.41 & 0.13\\
            -26.5 & 3.8 & 0.97 & 25.0719 & 0.0234 & 2.7770 & 0.0225 & 0.27 & 0.02 & -1.15 & 0.11 & -0.69 & 0.10 & 26.0 & 1.27 & 0.12\\
            -27.5 & 3.8 & 1.35 & 25.1459 & 0.0677 & 2.8481 & 0.0650 & 0.31 & 0.06 & -0.68 & 0.20 & -0.25 & 0.16 & 26.0 & 1.76 & 0.21\\

            \hline
        \end{tabular}
    \end{table}
\end{landscape}

\subsection{Radio flux density distribution for red quasars}

\begin{figure*}
        \includegraphics[width=\textwidth]{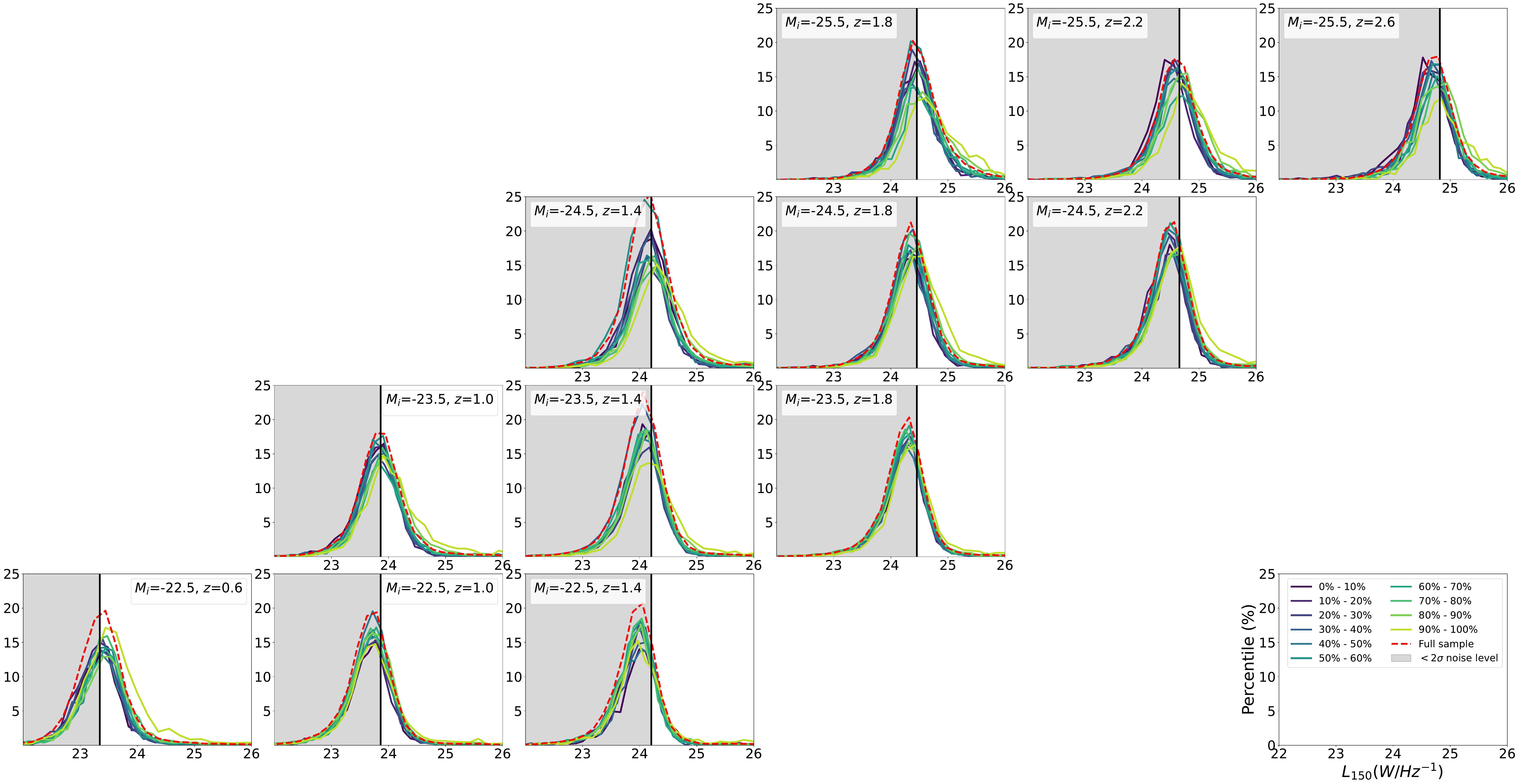}
        \caption{The radio luminosity distributions separated by 10-percentile colour sub-grids (solid lines) compared with the radio luminosity distribution of the entire population (red dashed lines), plotted for every $M_i-z$ grid investigated in Section~\ref{sec:colour}. Higher percentiles indicate higher values in $\Delta E(B-V)$ colour excess, hence redder quasars compared to the entire population. Across all $M_i-z$ grids, the distributions for rQSOs (top 10\% of the $\Delta E(B-V)$ colour excess) show an extended wing on the radio-bright end while sharing similar star-formation peaks when compared to the entire population. Note that the shaded areas reflect sources with flux densities below the 2-sigma LoTSS DR2 uncertainty limit and therefore cannot be used to visually determine the exact positions of the star-formation peak or the relative jet power unless fitted with our two-component model.}
        \label{fig:sample_dist_color_full}
\end{figure*}

In Figure~\ref{fig:sample_dist_color_full}, we present the actual radio luminosity distributions of different colour percentile quasars across all $M_i-z$ grids used for the study of colour dependence in Section~\ref{sec:colour} (note that sources with negative flux densities are excluded when plotting with radio luminosities). The radio luminosities are binned into the same bins defined for the top panel in Figure~\ref{fig:sample_dist_color}, and are plotted in the colour-code defined for different colour percentiles as shown in the bottom-right legend. The red dashed line shows the radio luminosity distribution for the entire population within the grid. Here we would like to remind the readers that a fair amount of sources lie under the $2\sigma$ flux density uncertainty limit; therefore the distributions presented in Figure~\ref{fig:sample_dist_color_full} cannot be used to deduce the exact levels of SF and AGN activity. Readers should still refer to Figure~\ref{fig:result_color_mcmc} for the model-fitted values of $L_\mu$ and $f$ in different colour-split samples. Across all $M_i-z$ grids, the radio flux density distributions for rQSOs (defined as the top 10\% in colour excess) show an extended radio-bright wing while sharing similar star-forming peaks when compared to the distributions for the entire population. This is in accordance with the upper panel in Figure~\ref{fig:sample_dist_color} and the best-fit parameter values presented in Figure~\ref{fig:result_color_mcmc}.

%\begin{landscape}
%\end{landscape}

%%%%%%%%%%%%%%%%%%%%%%%%%%%%%%%%%%%%%%%%%%%%%%%%%%

% Don't change these lines
\label{lastpage}
\end{document}